# Air temperature and humidity during the solar eclipses of 26 December 2019 and of 21 June 2020 in Saudi Arabia and in other eclipses with similar environments

Marcos A. Peñaloza-Murillo

*Universidad de los Andes, Facultad de Ciencias, Departamento de Física*
*Mérida, Edo. Mérida 5101, Venezuela*
&
*Department of Astronomy, Williams College*
*Williamstown, Massachusetts 01267. U.S.A.*
(map4@williams.edu)

Abouazza Elmhamdi

*Department of Physics and Astronomy, King Saud University*
*P. O. Box 2455. Riyadh 11451, Saudi Arabia*
(a.elmhamdi@gmail.com)

Jay M. Pasachoff

*Department of Astronomy, Williams College*
*Williamstown, Massachusetts 01267, U.S.A.*
(jmp@williams.edu)

Michael T. Roman

*The University of Leicester, Department of Physics and Astronomy*
*Leicester LE1 7RH, U.K.*
(michael.thomas.roman@gmail.com)

Yu Liu

*Yunnan Observatories, Chinese Academy of Sciences, Kunming 650216*
*People's Republic of China*
(lyu@ynao.ac.cn)

Z. A. Al-Mostafa

*King Abdulaziz City for Science & Technology, National Center for Astronomy*
*and Navigations, P.O. Box 6086, Riyadh 11442, Saudi Arabia*
(zalmostafa@kacst.edu.sa)

A. H. Maghrabi

*National Centre for Applied Physics, King Abdulaziz City for Science and Technology*
*Riyadh 11442, Saudi Arabia*
(amaghrabi@kacst.edu.sa)

H. A. Al-Trabulsy

*Department of Physics and Astronomy, King Saud University*
*P. O. Box 2455 . Riyadh 11451, Saudi Arabia*
(haltrabulsy@yahoo.com)

Abstract

Few investigations have been published on the atmospheric response to a solar eclipse in desert and similar environments. In order to contribute to the improvement of this field, we report air temperature and humidity changes during the two annular solar eclipses of 26 December 2019, and 21 June 2020, which we studied, respectively, from the cities of Al-Hofuf and Riyadh in the Kingdom of Saudi Arabia. During the December (annular) eclipse the Sun rose already eclipsed (~91.53% of the area covered) while the June eclipse, although also annular in other places of the Arabian Peninsula, was just partial at Riyadh (area covered ~ 72.80%), starting about two hours after sunrise. This difference apparently affected the observed response on the recorded variables of temperature, relative humidity (RH) and vapor pressure (VP) in different ways in the two events. Change in these variables went unnoticed for the first eclipse since it was within the natural variability of the day; yet for the other, they showed clearly some trend alterations, as a result of the eclipse, which we analyze and discuss. A decrease in air temperature of 3.2 °C was detected for the 73% in Riyadh; however, RH and VP showed an oscillation that we explain in the light of a similar effect reported in other eclipses. We found a time lag of about 15 min measured from the eclipse central phase in this city. We made a meaningful inspection of related fluctuations and dynamics from the computed rates of the temporal variation of temperature and RH. Trying to identify and understand the influence of solar eclipses in similar environments we have made a broad inter-comparison with other observations of these variables in the Near East, northern Africa and in the United States. Additionally, we compare our results with results obtained by other authors working with the December eclipse but in the United Arab Emirates and Oman, which showed dissimilar results. These inter-comparisons show to some extent how effectively the lower atmosphere can respond to a solar eclipse within a desert environment and others similar. At the end, and in contrast, we suggest similar inter-comparative studies to investigate the atmospheric response to a solar eclipse under extreme environmental conditions prevailing in dry locations comparable with those found in or near polar zones. As a preamble, and for completion of the paper, a historical revision of temperature and humidity in the context of eclipse meteorology is also included.

## 1. Introduction

Since ancient times solar eclipses during which the Moon obscures fully or partially the Sun's disk, have been considered to be fascinating astronomical phenomena (Olson & Pasachoff, 2019). Among the different types, annular eclipses –with an annulus of solar photosphere remaining visible at maximum coverage- occur with about the same frequency as total solar eclipses about every 18 months. During this century, Espenak (2016) shows that 32.1% of eclipses are annular, 30.4% are total, and 3% are hybrid (with some annularity and some totality); for online maps, see Jubier (2020a) and [http://eclipses.info]. Annular eclipses occur when the Moon covers the center of the Sun's disk but because of its elliptical orbit is too small in angle to cover the entire solar photosphere (Pasachoff & Filippenko, 2019), producing a magnificent annulus (ring-shape). Unfortunately, the misleading term "ring of fire" has often appeared in the press in recent decades instead of, for example, "ring of photosphere."

A chronology of ancient solar eclipses given by Schove & Fletcher (1987) between 1 ∼ to 1000 CE indicates that some of them were observed from the Arab world, for example, those occurred on 19 May 486 (possibly in Syria or Arabia), 1 August 566 (total in Arabia), 27 January 632 (first Islamic solar eclipse), and 7 December 671 (in Syria and Arabia). In the interval of 977-981 and in the years of 983 and of 985-986 Cairo, the Egypt's Capital, witnessed solar eclipses [from which two total (2-3 May 980 and 1-2 March 983) and three annular (8 June 978, 14-15 May 979 and 28 May 979 were seen in that country)]. Others were observed from Iran (360, 484, 873 and 893) and Iraq (604, 693 and 969), etc. Prior to our common era (CE) Finnegan (1998) reports the observation of an eclipse in Assyria on 15 June 763 and of another two in Babylon, 22 April 621 and 4 July 568. On the other hand, Khalisi (2020) reports three annular eclipses in 1399, 1389 and 1378 observed in ancient Egypt. More information on solar eclipses in the ancient Near-East between 3000 BCE to 0 can be found in Kudlek & Mickler (1971); another sources are Stephenson (1975) and Ben-Menahem (1992). For medieval Arabic eclipse observations see Said & Stephenson (1991).

From 2000 up to now there have been, among partial, annular and total, forty-nine solar eclipses around the world (Espenak, 1987, 2016). Some of them have been observed from countries of northern Africa and from the Near-East. Measurements of the atmospheric response during the annular solar eclipse of 3 October 2005 have been reported in these regions by Hassan *et al*. (2010) in Egypt and by Anbar (2006) in Saudi Arabia. Also during the total solar eclipse of 29 March 2006, measurements made by Uddin *et al*. (2007), Pintér *et al*. (2008), Pleijel (2008) and Stoev *et al*. (2008, 2012) in Turkey, by Nawar *et al*. (2007) in Egypt, by Hassan & Rahoma (2010) in Libya and, by Nymphas *et al*. (2009) in Nigeria, were reported. Recently the solar eclipse of 1[st] September 2016 was observed by Ojobo *et al.* (2017) in Nigeria and the annular solar eclipse of 26 December 2019 was observed by Nelli *et al*. (2020) in the United Arab Emirates (U.A.E.) and Oman. Special emphasis was put in these investigations in observing atmospheric physical variables like sky brightness, solar and net radiation, air temperature, humidity, wind, etc.

The most recent solar eclipse visible from the Near East was observed in Saudi Arabia on 21 June 2020 as partial [views of annularity from other countries appear in Pasachoff, (2021)]. In particular, measurements of air temperature and relative humidity (RH) were made in the city of Riyadh. In addition, similar measurements were made during the prior annual solar eclipse of 26 December 2019 in the city of Al-Hofuf (Al-Ahsa region) (Liu *et al.*, 2020). In this paper we report these measurements and we present here an analysis based on a comparison between both set of observations taking into account that these eclipses, as partial ones, occurred in the Arabian Peninsula, which is characterized for being a desert and arid region, but noting that the first one (annular) was in December close to the boreal winter solstice, at the sunrise, and the other (73% at max) just on the boreal summer solstice also in the morning; in other words, the two eclipses have, among other features, a time interval difference of 6 months, occurring hence in two different seasons. Although both eclipses were in the morning, one early during sunrise and the other about 2 hr 5 min later, the effects on air temperature and humidity of the first one were not as severe as for the second one. We found that the local weather and geographical conditions of the two eclipse observation sites were different such that it had an impact on the air temperature and humidity overall results. Considering that other authors have published results on atmospheric effects of the recent annular eclipse of 2019 in the Arabian Peninsula but in Abu-Dhabi (United Arab Emirates) and Oman (Nelli *et al.*, 2020) we will try, additionally, to make a comparison as far as possible with their observations. These solar eclipses are not the first observed scientifically from the Near East or even in the region of the southern Mediterranean basin (northern Africa); earlier atmospheric observations were carried out by Jaubert (1906) during the total solar eclipse of 30 August 1905 in Algeria (Constantine Region), by Klein & Robinson (1955) during the partial solar eclipse of 25 February 1952 in many sites of Israel and, by Rahoma *et al.* (1999) and Hassan *et al.* (1999) during the partial solar eclipse of 11 August 1999 in Egypt. More recently, Anbar (2006) working with a partial eclipse (60%) which took place over the Hada Al-Sham area, Makkah region, studied atmospheric effects on the air layer near the surface during the solar eclipse of 3 October 2005; some comparison with our results are in order.

## 2. Eclipse meteorology in the past and now: temperature and humidity

### 2.1. Temperature historical overview

The earliest recorded observations ever made of air temperature during a solar eclipse corresponds to Brereton (1834) during the solar eclipse of 30[th] November 1834 at Boston, Massachusetts, and from that time onwards the era of instrumental eclipse meteorology began [see Barlow (1927) for an extensive bibliography (in different languages) up to that year]. In the nineteen-century, between that year and 1899, at least thirteen studies or reports were published covering air temperature measurements made during the observation of nine solar eclipses spread over the U.S.A., England, Russia, India, and the Caroline Islands (Pacific Ocean). Next, at least ninety studies and reports were published between 1900 and 1999 containing temperature measurements carried out during the observation of thirty-six solar eclipses around the world. It amounts about to 1.82 publications each two years on average. During this interval, the eclipses studied most thoroughly were that of 16 February 1980, which took place over India, with eleven studies, and that of 11 July 1991, which took place over Mexico, Central America, and Colombia with seven studies. From 2000 to date, among studies and reports, at least eighty-eight papers have been published related to thirteen solar eclipses (about 6.77 papers per eclipse) with the eclipse of 11 August 1999 mostly over Central Europe that received the most coverage with twenty-one publications followed by the eclipse of 20 March 2015 with fourteen papers (e.g., Harrison & Hanna, 2016); these totals imply a rate of 4.4 publications per year up to 2020 since 2000 (20-year span), more than the interval rate of 1900-1999 (100-year span). Aplin *et al.* (2016) review some of these articles and Peñaloza-Murillo & Pasachoff (2015)'s supplementary material compiles an extensive but non-exhaustible account of this type of studies between 1842 and 2013 [see also Kameda *et al*. (2009), table 1]. In this way, undoubtedly, the fall of temperature in the free air near the Earth's surface has become, and still is, the most marked, studied and unquestioned response of the atmosphere observed during solar eclipses (along with solar radiation whose fall currently precede the temperature) in history up to the present (Nelli *et al*., 2020; Pakkattil *et al*., 2020; Reddy T. *et al*., 2020). It is conclusive that, by and large, all these studies have shown that a temperature drop (or anomalies) and lag vary in wide magnitudes and fashions depending on the local environmental, meteorological and astronomical circumstances.

## 2.2. Humidity: previous studies and rationale

Regarding humidity observations, Ahrens *et al*. (2001) and Pleijel (2008), states that the literature contains fewer observations of humidity time series during solar eclipses than of temperature series. This conclusion is shared by Bhat & Jagannathan (2012) who states that humidity has not been given importance beyond simple reports of variation in the relative humidity (RH) or water vapor pressure (VP). However, that is not true. Early and historical measurements of humidity changes during solar eclipses began in the eighteenth century when an annular solar eclipse was observed in the continental United States on 26 May 1854. These first historical observations of humidity, through measurements of dew point (DP), were performed by Alexander (1854). Four years later a total solar eclipse was also taking place in the U.S. on 18 July 1860 and some heavy dew was reported by Gilliss (1861) during that event; a light fog formed at sunrise and heavy dew continued to accumulate. Subsequently, Upton & Rotch (1888, 1893) undertook the first measurements of RH during the total solar eclipse of 19 August 1887 at Chlamostino (Russia) and, during the total solar eclipse of 1 January 1889 at Willows (Colusa County), California. In the latter eclipse, dew formed on grass and other exposed surfaces; sheets laid out to facilitate the detection of shadow bands became very moist and the authors stated that the probable slight increase in vapor pressure after totality likely contributed to the rise in RH.

At the onset of the twenty-century, Clayton (1901) offered some preliminary explanations to account for the variation observed in VP. Following the TSE of 28 May 1900, Bigelow (1902) computed VP from observations of 61 stations throughout southern U.S. States and, based on the mean curve over these stations, found a decrease at the time of the minimum cooling.  Later on, in a compilation from numerous measurements of humidity during solar eclipses were made by Clayton (1901) and Bigelow at eclipses [1883 (Caroline Island, Pacific Ocean), 1887 (Russia), 1889 (California, USA), 1898 (India), 1900 (Southern States, USA), 1901 (Sumatra) and 1905 (Atlantic Ocean, Spain, Algeria), Clayton (1908) found a maximum VP [i.e., absolute humidity (AH)] 30 to 50 min *preceding* totality in addition to a minimum at the observation nearest totality as well as a second maximum after totality. Since this pattern was observed not only on the ground but also in the free air at 300 m, Clayton ascribed the minimum to the descent of drier air at the time of totality. Measurements of RH and/or VP continued throughout the following decades, including work from the following: Bauer & Fisk (1916) who published data taken at the Ekaterinburg Observatory (Russia) by H. Abels and P. Mueller during the solar eclipse of 21 August 1914; of Ugueto (1916) and Sifontes (1920) during the total solar eclipse of 3 February 1916 and the partial solar eclipse of 22 November 1919 both in Venezuela; of Bilham (1921) who published data taken by R. Francis Granger at Lenton Fields Climatological Station in Nottingham, England, during annular solar eclipse of 8 April 1921; of Samuels (1925) observing from Texas and North Dakota in the United States the solar eclipse of 24 January 1925; of Stenz (1929) who went to Jokkmokk, Sweden, to observe the solar eclipse of 29 June 1927; of Stratton (1932) who observed from Canada the solar eclipse of 31 August 1932; of Cohn (1938) who published data of both the solar eclipse of 31 August 1932 in Maine (in the northeastern United States), and the solar eclipse of 14 February 1934 in east Caroline Islands, south

Pacific Ocean, and of Brooks *et al*. (1941) across New England (northeastern United States), during the solar eclipse of 31 August 1932.

An attempt to explain the coincidence of the minimum of VP with the time of totality was offered initially by Süring *et al*. (1934) [cited by Brooks *et al*. (1941)] where both AH, and therefore the VP, depends essentially on the evaporation from the surface of the earth or whatever surface give moisture to the air (e.g., vegetation). The temperature of these surfaces follows directly from the heat balance, so the evaporation and therefore the VP must follow closely the solar radiation. Thus, with this radiation diminishing, the amount of water evaporated decreases (Trapasso & Kinkel, 1984) and may not be sufficient to replace the moisture still transported even by the reduced convection (Peñaloza-Murillo & Pasachoff, 2015; Peñaloza-Murillo *et al*., 2020) to higher regions by mass exchange. Furthermore, a settling and mixing of drier air from aloft may help reduce the moisture. Only with increasing evaporation after totality does the VP rise again. The reader who wants to explore in greater detail many of the issues, related to convection decay during a solar eclipse, is referred to the early works of Antonia *et al*. (1979), Raman *et al* (1990), and Eaton *et al*. (1997) and subsequently works of Dolas *et al*. (2002), Szałowski (2002) and Anfossi *et al*. (2004). This effect seems to be confirmed by the fact that in the central area of the eclipse there is a distinct minimum of vapor pressure coinciding closely with the rise of barometric pressure as it is illustrated in Figure 1, thus indicating that there is a descent of drier air from aloft to supply the outflowing current. This decrease in VP occurred not merely at the ground, but also on high buildings and also was noted in a balloon and in a record by a kite, at an altitude of 200 meters in the eclipse of 1905. More evidences presented by Clayton (1908) at that time showed that it was a descending current set up by the chilling of the atmosphere during the passage of the Moon's shadow, and not the radiation from the ground, the cause of the minimum of VP, which is nearly coincident with totality and with the time of maximum cooling of the air away from the Earth's surface, as measured at the kite.

During the annular solar eclipse of 30 May 1984, visible across southeastern United States, and as a partial nationwide, Trapasso & Kinkel (1984) measured the evaporation rate during the eclipse. Effectively, they found that a drop in temperature, a rise in RH, and a drop in wind speed retarded the evaporation process. The evaporation recorder they used at the College Heights Weather Station on the Western Kentucky University campus showed that before the eclipse (8:00-9:00, CST), the evaporation rate was approximately 0.45 kg-m$^{-2}$-h$^{-1}$. During the maximum of the eclipse (10:00-11:00, CST), the evaporation rate leveled off to 0.20 kg-m$^{-2}$-h$^{-1}$ while one hour after the eclipse (12:00-13:00, CST), this rate rose to a value of 0.75 kg-m$^{-2}$-h$^{-1}$. On the other hand, Bose *et al*. (1997) found during the deep morning partial solar eclipse of 24 October 1995, in Delhi, a decrease in water vapor column (mg-cm$^{-2}$) as a function of solar zenith angle in relation to days before and after the eclipse.

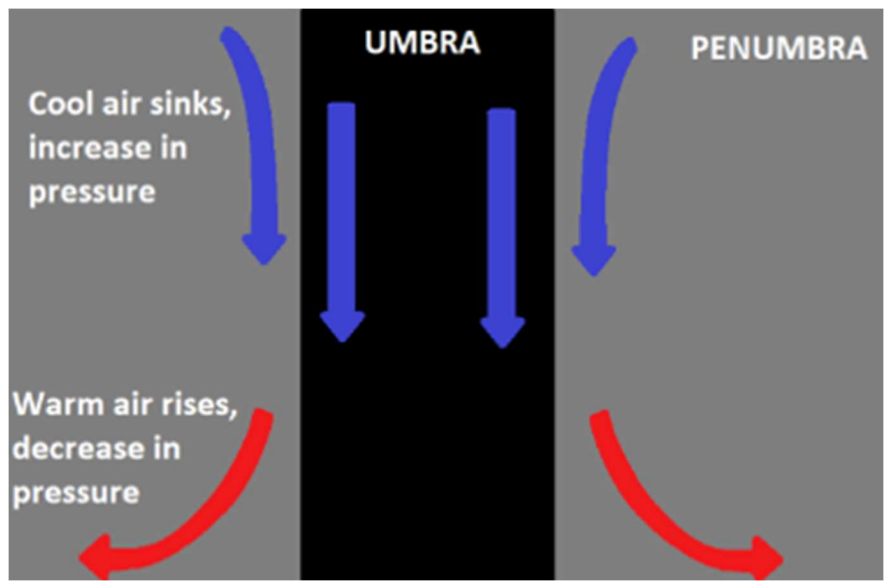

**Figure. 1**. In the central area of the eclipse there is a distinct minimum of vapor pressure coinciding closely with the rise of pressure, which is nearly coincident with totality and with the time of maximum cooling of the air away from the Earth's surface; this indicates that there is a descent of drier air from aloft to supply the outflowing current as it is illustrated in this figure. It is possible, then, that this produces some adiabatic warming at the surface, partly mitigating at least the decrease in temperature due to the loss of radiative heating. Yet, presumably, the radiative effect dominates.

Clayton (1908) suggested that this descent by virtue of a direct cooling of the air, leading to a fall of VP, was as large over the ocean as it was over land areas, where the fall of surface air temperature was much larger, but where friction retarding air movement was also much greater. The VP falls as the result of descending air because, as a rule, it decreases with height and a descending air like that brings about a fall in VP at lower levels. On the contrary, however, it has been noted that sometimes there has been an increase in VP instead of a decrease, which Clayton (1908) attributed to an overflowing air current with a higher VP than the surface current as it was shown by observations with kites signifying that if an eclipse were to occur under such conditions, a descending current of air in the central area of the eclipse would bring about a rise of VP at lower levels; this would explain observations such as those of Upton & Rotch (1892) at Willows, California, and of Eliot (1899) in Nagpur, India. In the latter case Brooks *et al*. (1941) stated that "… The marked decrease in wind velocity during the eclipse [22 January 1898] would reduce the rate of evaporation from the wet bulb, and hence the difference in reading between the two [thermometers], producing an artificially apparent increase in vapor pressure." Years later these authors working with the eclipse of 31 August 1932, in northeastern United States (New England) obtained widely varying results (curves) from observations of 300 stations. Looking at their tables 20 and 21, we noted that there was no "orderly sequence in the amount of the vapor pressure change, and that, although the resulting average is usually plus, hardly more than half the stations show a plus departure."

Further analysis by Brooks *et al*. (1941) discusses the VP changes during the eclipses of 1932 (afternoon eclipse) and 1900 (morning eclipse) both in the United States, to get more insight on these changes. From the 300 stations used for the first eclipse and the 73 stations used for the other, changes were observed of the eclipse dividing its two-hour time into four quarters of a half an hour each. The trend of VP was tabulated as positive (or increasing), no change, or negative (or decreasing) for each quarter. From their table 22 it was observed a tendency of the VP to rise in the third quarter just after totality (especially in zones of 80-89% occultation and totality). Also this table shows positive trends to be generally more frequent than negative in the second quarter. In the case of the 1900 eclipse Bigelow (1902)'s data detected a VP fall at ~ 15 min preceding the maximum eclipse, with a gradual rise of the same amount in the following hour.

It is difficult to obtain a clear picture of any typical effect, wrote Brooks *et al*. (1941) at that time, but perhaps –they added- some explanations of apparent inconsistences may be offered. For the first time in history of eclipse meteorology an attempt to explain the response of VP to solar eclipses was offered.

To begin with, they refer to two main factors involved in any change of VP in the lower portion of the atmosphere in fair weather: the rate of evaporation from the surface, and the rate of turbulent mixing of the humidity surface air with the general air. To quantify the latter in fair weather, a diffusion coefficient for turbulent transfer of water vapor in the air has to be calculated, which depends on von Karman's constant, friction velocity, mass concentration of water vapor, and height above surface (see WMO, 1958: 19-36; Lowry & Lowry, 1989; Monteith & Unsworth, 1990: 234-239); thus, values of the rate are very variable under normal conditions. In a solar-eclipse situation, as long as this develops, evaporation from the ground is considerably reduced. Additionally, a reduction in wind speed occasioned by the eclipse would also favor a decrease in net evaporation -although this is not its main effect- and an accentuation of local difference. As a result, a slight decrease in VP is expected during the first half of an eclipse, accentuated by condensation in cases where ground surface cools to the dew point (DP) (although this explanation fails to explain situations where observations are made at stations on tall buildings in cities as there is little or no evaporation taking place from dry streets and buildings along with any change in evaporation from the surroundings countryside which could hardly be felt in a measurable degree). At stations near water, a slight shift in the wind direction might cause a larger change in VP than that attributable to the eclipse, which over water would be very small owing to the small change in surface temperature.

The second factor, the most important factor controlling VP according to Brooks *et al*. (1941), is the rate of turbulent mixing, over which an eclipse may has an effect. Considering that a source of water vapor is the Earth's surface if it is not carried aloft then VP increases, and this is typical of late afternoon and early evening. A rise in AH was observed during the half hour after totality of the afternoon 1932 eclipse, which was definitively related to the changes in VP (Brooks *et al*., 1941). The reason for that was a marked decrease in convection, which reached a minimum after totality (Antonia *et al*., 1979; Raman *et al*., 1990; Eaton *et al*., 1997; Dolas *et al*., 2002; Szałowski, 2002; Anfossi *et al*., 2004; Peñaloza-Murillo *et al.*, 2020). With returning sunshine, though, increasing evaporation from the immediately warming surface may be also a factor in the open country. Accordingly, the convection factor was the cause of a larger increase of VP in the eclipse of 1932 than in that of 1900 (which occurred at 10:00 near the maximum of VP, before day-time convection is well stablished). To conclude their analysis, Brooks *et al*. (1941) state that the net change in VP due to an eclipse is relatively small and quite diverse, being the result of opposing factors, evaporation and convection (though aerological data seem to minimize this effect). Summing up, in the first half of an eclipse in clear weather, the VP is likely to fall, apparently because the rate at which the air is charged by evaporation owing to decreasing convective exchange, whereas immediately after totality, however, the VP usually rises, probably because evaporation is then increasing while exchange is still decreasing.

As to RH Clayton (1908) explains it is determined by two factors acting in opposite directions: the chilling of the air tending to increase RH and a diminution of the VP, tending to lower the RH. Over land areas, where the fall of temperature is large, the first effect usually predominates, and the RH shows a maximum at the time of minimum temperature. However, Clayton (1908) reported a case where the effects were so nearly equal that the RH curve showed only a series of irregular oscillations during the eclipse of 1905 in Burgos, Spain. Moreover, in the same eclipse but at the kite, 300 m above sea level (asl), showed for the same eclipse a marked fall in RH, notwithstanding the fall in temperature, which strongly confirms the conclusion that there is a descending current in the middle of the eclipse area (Figure 1), for, as is well known, descending currents are dry owing to the dynamics warming of the air without increase of AH. Something similar occurred with the partial solar eclipse of 30 May 1984 observed by Knöfel (1986) in Germany from five different places, where RH during the eclipse had an irregular behavior than the obvious rise of it.

After the pioneering work of Brooks *et al*. (1941) giving analysis and explanations for VP, AH and/or RH changes during solar eclipses, these variables continued being measured along the 20[th] century in different solar eclipses worldwide; for example, in Japan during the total solar eclipse of 9 May 1948 (Huzimura, 1949; Ushiyama *et al*., 1949); in Israel during the partial solar eclipse of 25 February 1952 (Klein & Robinson, 1955); in Scandinavia during the total solar eclipse of 30 June 1954 (Kullemberg, 1955; Paulsen, 1955); in Yugoslavia in the eclipse of 7 March 1961 (Anić, 1970); in California in the eclipse of 20 July 1963 (Pruitt *et al*., 1965); in Canada in the eclipse of 10 July 1972 (Stewart & Rouse, 1974); in Greece in the eclipse of 29 April 1976; (Katsoulis, 1976); in India in the eclipse of 16 February 1980 (Babu & Sastry, 1982; Mohanakumar & Devanarayanam, 1982); in the United States in the eclipse of 30 May 1984 (Trapasso & Kinkel, 1984) and Germany (Knöfel, 1986); in Costa Rica (Fernández *et al*., 1993), Mexico (Bernard *et al*., 1992; Gaso *et al*., 1994) and California (Mauder *et al*., 2007) during the eclipse of 11 July 1991; in Paraguay in the eclipse of 3 November 1994 (Fernández *et al*., 1996);  in India again  during the solar eclipse of 24 October 1995 (Arulraj *et al*., 1998; Bansal & Verma, 1998; Bose *et al*., 1997; Dani & Devara, 2002; Ghosh *et al*., 1997; Gonzalez, 1997; Jain *et al*., 1997;  Niranjan & Thulasiraman, 1998; Sapra *et al*., 1997; Singh *et al*., 1999); in Venezuela during the total solar eclipse of 26 February 1998 (Peñaloza-Murillo, 2002, 2003); and during the last solar eclipse of the twenty-century and the Second Millennium on 11 August 1999, over Europe (Ahrens *et al*., 2001; Božić *et al*., 2003; Crochard & Renaut, 1999; Foken *et al*., 2001; Häberle *et al*., 2001; Kolarž *et al*., 2005;  Kolev *et al*., 2005;  Perkins, 2000; Prenosil, 2000;  Simeonov *et al*., 2002; Stoev *et al*.,  2000; Tzenkova *et al*., 2002), over Egypt (Rahoma *et al*., 1999), and over India (Krishnan *et al*., 2004)

The first total solar eclipse of the twenty first-century and the Third Millennium took place on 21 June 2001, over southern Africa and Madagascar. The only work that we found with RH measurements, taken in Zambia, for that eclipse is a report of Imbres (2001) which is archived at the U.S. Coast Guard Academy´s Center for Advanced Studies (New London, Connecticut). It was not until the solar eclipse of 3 October 2005, over Saudi Arabia when Anbar (2007) reported the next RH measurements. Subsequently, the total solar eclipse of 29 March 2006, brought about a series of papers containing RH and/or VP measurements made in different countries as, for example, in Lybia (Hassan & Rahoma, 2010), in Turkey (Pleijel, 2008; Stoev *et al*., 2008; Uddin *et al*., 2007), in Greece (Amiridis *et al*., 2007; Tzanis *et al*., 2008) and in Russia (Kadygrov *et al*., 2013). In the Svalbard Archipelago in Norway, Sjöblom (2010) made RH observations during the deep partial solar eclipse of 1 August 2008, and in Novosibirsk, Russia, where it was total, Kadygrov *et al*. (2013) made observations of water content.

During the longest total solar eclipse of the twenty-one century in the Far East on 22 July 2009, some observations were made in different countries, such as in South Korea (Chung *et al*., 2010; Jeon, 2011), in China (Chen *et al*., 2011; Lu *et al*., 2011; Pintér *et al*., 2010; Stoeva *et al*., 2009; Zainuddin *et al*., 2013) and in India (Rao *et al*., 2013; Kumar, 2014). On 15 January 2010, another solar eclipse (annular) was visible in India and some observations in this country were reported in the literature (Babu *et al*., 2011; Bhat & Jagannathan, 2012; Manchanda *et al*., 2012; Muraleedheran *et al.*, 2011; Subrahamanyam *et al*., 2011; Subrahamanyam & Anurose, 2011; Subrahamanyam *et al*., 2012; Vyas *et al.* 2012). Later, in the partial eclipse of 4 January 2011, over Moscow, Kadygrov *et al.* (2013) measured the water content. After that, came the total solar eclipse of 20 March 2015, for which some publications were made reporting measurements in different regions of Europe, for example, in Czech Republic (Nezval & Pavelka, 2017), in southern Italy (Romano *et al.*, 2017), in the United Kingdom (Burt, 2016), and in France (Kastendeuch *et al.*, 2016). In recent times Ojobo *et al.* (2017) presented RH measurements made in Nigeria during the annular solar eclipse of 1 September 2016, and Paramitha *et al.* (2017) did the same but for the previous partial eclipse of 9 March 2016, in Indonesia. Lately, during the total solar eclipse of 21 August 2017, in the United States meteorological networks across the country monitored widely humidity variables along and perpendicular to the shadow track as well as in the penumbral zone (Buban *et al.*, 2019; Burt, 2018; Lee *et al.*, 2018; Mahmood *et al.*, 2020; Turner *et al.*, 2018).

*2.3. Reviewing*

Summing up, relativity humidity of the air is an atmospheric or meteorological variable that, from time to time, is measured during solar eclipse. Along with this variable, occasionally absolute humidity (vapor pressure) and water vapor (mg-cm$^{-2}$) changes at the surface also have been measured. In comparison to air temperature, the literature contains fewer observations of air humidity series under particular circumstances.

Reviewing some of the results of the previously cited references, it can be said that early measurements of changes in absolute humidity due to eclipses of 1898, 1900 and 1932 were made in the U.S. (Bigelow, 1902) as commented upon by Brooks *et al.* (1941). And during the 1932 TSE, these authors performed RH measurements at the Earth's surface. For the eclipse of February 3, 1916, over Venezuela, Ugueto (1916) recorded data of air humidity. For a long time afterwards, during the TSE of 25 February 1952, in Africa and Asia (Near, Middle and Far East), Klein & Robinson (1955) made measurements of RH and water pressure at Eilath, Israel, where it was visible partially.

Modern observations of water vapor were reported by Bose *et al.* (1997), who found that it had decreased during the 95.7% maximum phase of the solar eclipse of 1995 over Delhi. At this same event, Gonzalez (1997) found that RH increased to 76% near mid-eclipse, which was 32% higher than the previous morning at Neem Ka Thana (India); also, Jain *et al.* (1997) found that total water content varied roughly from 0.5 gm-cm$^{-2}$ to 1.2 gm-cm$^{-2}$ during the period of this eclipse at the same place, giving a variation of approximately 8 min to 7 min in water vapor. Kolarž *et al.* (2005) reported a "mirror symmetry" between temperature and RH during the eclipse of 11 August 1999, at Belgrade (maximum obscuration 97.7%), reflecting the strong dependence of RH on temperature. Ahrens *et al.* (2001) for this eclipse showed that weather conditions were not ideal in southwest Germany to investigate impacts on air humidity; that should be the major reason why no typical pattern of water pressure was observed at the Plittdersdorf site during the eclipse. Moreover, the fluctuation (increase and decrease) of the RH was caused by the behavior of temperature and is, therefore, not occasioned by the water vapor content in the lower planetary boundary layer (or atmospheric boundary layer, ABL).

Not long ago, Pleijel (2008) showed an inverse pattern of RH compared to air temperature during the TSE of 29 March 2006, at Side, Turkey; in contrast, the vapor pressure varied little and did not appear to be systematically affected by the phenomenon. For this eclipse, observed by Kadygrov *et al.* (2013) in Novosibirsk, Russia, the minimum of total water vapor content of about 0.65 g-cm$^{-2}$ was observed $\sim$ 40 min before the total eclipse phase. On the day of that eclipse, starting from 14:00 local time until the total phase of the solar eclipse, there were considerable turbulent water vapor quasi-periodic pulsations as large as from 0.1 to 0.25 g-cm$^{-2}$, or more than 8%. Usually, the value of this fluctuation is lower than 0.02 g-cm$^{-2}$, which is $\sim$1%. The amplitude of these pulsations monotonically decreased during and after the eclipse, correlating well with the decrease of the temperature (and with the decrease of the pulsations of the vertical wind velocity component. This indicates that the convection was weakened not only in the near ground layer, but also in the ABL up to about 2-km height (characteristic altitude of water vapor). The observed quasi-periodic pulsations of water vapor content indirectly indicate the possible formation of a vertical circulation cell in the shadow area. The much moister air was removed from this part of the ABL due to increased pressure, and dry air entered from the middle troposphere. Such a type of circulation may give rise to internal atmospheric waves, which, in turn, cause quasi-periodic fluctuations in water vapor content in an atmospheric column.

When the annular solar eclipse of 15 January 2010, passed over southern India, Bhat & Jagannathan (2012) measured the moisture depletion in the surface layer at a couple of locations: Ramanathapuram and Mandapam (Tamilnadu). They found that during the entire 11-day study period, the lowest value in RH occurred towards the end of the eclipse. Specific humidity decreased by 2 g-kg$^{-1}$ during the eclipse and continued to decrease for few more hours. It recovered to the pre-eclipse values in the following afternoon. Humidity decrease was attributed to increased subsidence of drier air during and after the eclipse.

In the American eclipse of 2017, the Kentucky Mesonet (a meso-meteorological network) detected a steady decline of RH as the day progressed, followed by a sharp increase of nearly 40% (from ~40% to ~80%) during totality (about noontime), and a subsequent decline after the end of totality. With the commencement of the eclipse and the absence of solar forcing, air temperature declined and RH steadily increased. The latter decreased from its peak of about 75% in the early morning to near 40% prior to the beginning of the eclipse. During totality RH rapidly increased to about 60%; after the end it again decreased to about 42%. Subsequently, RH slowly increased, following its diurnal cycle. Regionally, RH also showed a spatiotemporal pattern through the eclipse evolution reflecting proximity to the path of the totality (Mahmood *et al.*, 2020). In Tennessee, Buban *et al*. (2019) found that measurements of water vapor mixing ratio showed a very gradual drying of the ABL aloft throughout the event; however, near the surface there was a slight increase in moisture just after totality ended, before drying continues

   After this brief bibliographical review and under the aforementioned considerations and contextual background let us see what happened with air temperature, RH and VP during the solar eclipse of 26 December 2019, and of 21 June 2020, in two places of Saudi Arabia during the morning of those days along with other solar eclipses observed under similar environments for comparison purposes.

## 3. Circumstances of the eclipses

### 3.1. Astronomical

   Both eclipses were followed and imaged during their different phases. Table I summarizes its main characterizing astronomical circumstances. In Figure 2 we highlight the visibility maps of the two events, with the bold line marking the center of the path and where the annular eclipse lasted longest. Each shaded area refers to the annularity location, with the eclipse duration gets shorter as we move closer to the edges [generated from Xavier M. Jubier's eclipse website (Jubier, 2020b)]. The solar eclipse of 26 December 2019, had a path of annularity of ~160 km width across an east-southern strip in Saudi Arabia, while the solar eclipse of 21 June 2020, created a narrower path of annularity only ~ 40 km in width across the extreme south of the country. The locations and coordinates of the cities, sites of our follow-up are also indicated in Figure 2, namely Al-Hofuf city and Riyadh city. In the bottom of Figure 2 we report a sample of our observations covering the main phases of both eclipses, where maxima phases are also highlighted. Note the magnificent and impressive annulus for the Al-Hofuf eclipse, and the at-most crescent partial eclipse solar shape for Riyadh. In Figure 3 and Figure 4 maps of the respective penumbral tracks are shown. Worth noticing is that some places in the Sultanate of Oman had two annular eclipses within a brief interval of six months approximately. One of us (JMP) observed the 2019 annularity from a site near the Kodaikanal Solar Observatory, south India.

**Table I**. Astronomical circumstances of the solar eclipses of 26 December 2019 and of 21 June 2020 in Saudi Arabia

| | **Annular solar eclipse** | **Partial solar eclipse** |
|---|---|---|
| **Date** | 26 December 2019 (D) | 21 June 2020 (J) |
| **Observing site & location** | Site: Al-Hofuf City (Four Mountains Camp) <br> -------------------------------------- <br> Location: 25° 17' N, 49° 42' E | Site: Riyadh City <br> -------------------------------------- <br> Location; 24° 44' N, 46° 37' E |
| **Timing sequence** <br> (Local time: UT + h) | Start of partial eclipse [$1^{st}$ contact (1CD)]: 05 hr 32 min <br> -------------------------------------- <br> Sunrise (partial eclipse in progress): 06 hr 28 min <br> -------------------------------------- <br> Annular eclipse starts [$2^{nd}$ contact (2CD)] at: 06 hr 34 min <br> -------------------------------------- <br> Maximum annular eclipse: 06 hr 36 min <br> -------------------------------------- <br> End of annular phase [$3^{rd}$ contact (3CD)]: 06 hr 37 min <br> End of partial eclipse [$4^{th}$ contact (4CD)]: 07 hr 48 min | Sunrise at: 05 hr 05 min <br> -------------------------------------- <br> Start of partial eclipse [$1^{st}$ contact (1CJ)]: 07 hr 10 min <br> -------------------------------------- <br> Maximum eclipse at: 08 hr 23 min <br> -------------------------------------- <br> End of partial eclipse [$4^{th}$ contact (4CJ)]: 09 hr 49 min |
| **Duration** | ~2hr 16 min (total) <br> -------------------------------------- <br> 2 min 59.2 s (with lunar limb corrected; annular phase) | ~2 hr 39 min (penumbral duration) |
| **[Moon/Sun] size ratio (at the eclipse maximum)** | ~ 0.956 | ~ 0.988 |
| **Obscuration (%)** | ~91.53 | ~ 72.80 |

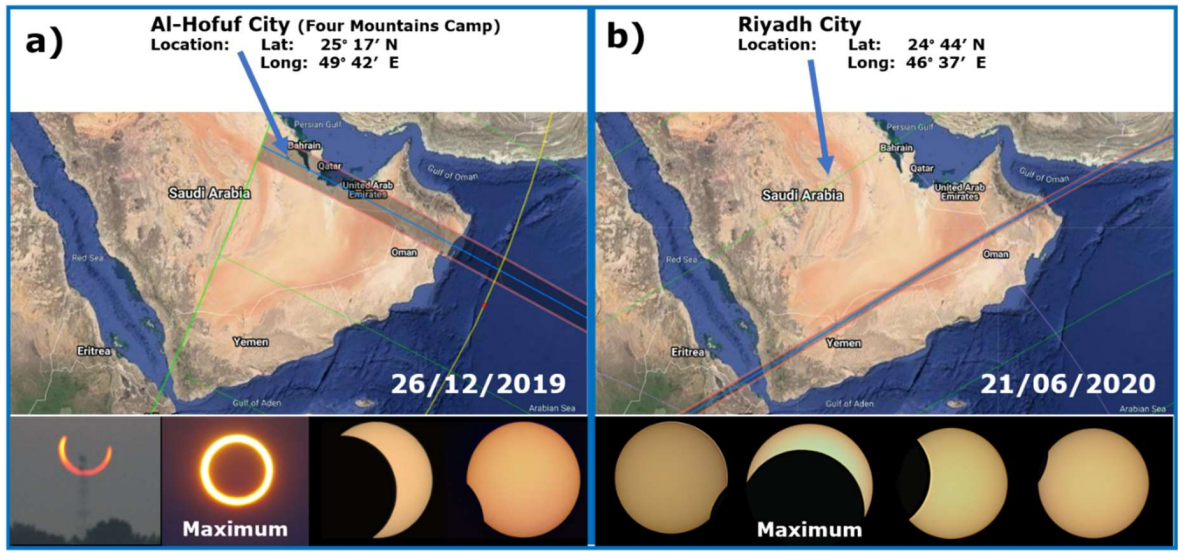

**Figure 2**. Upper panels: locations of the sites of the observations on the map of kingdom of Saudi Arabia at Al-Hofuf City [panel (a)] and Riyadh city [panel (b)]; the paths where the eclipse was seen as annular are also shown (as strips).  Bottom panels: a time sequence sample of our observations of the two eclipses. Images around maximum phases are highlighted. See Figures 2 and 3 for more details on the penumbral tracks.

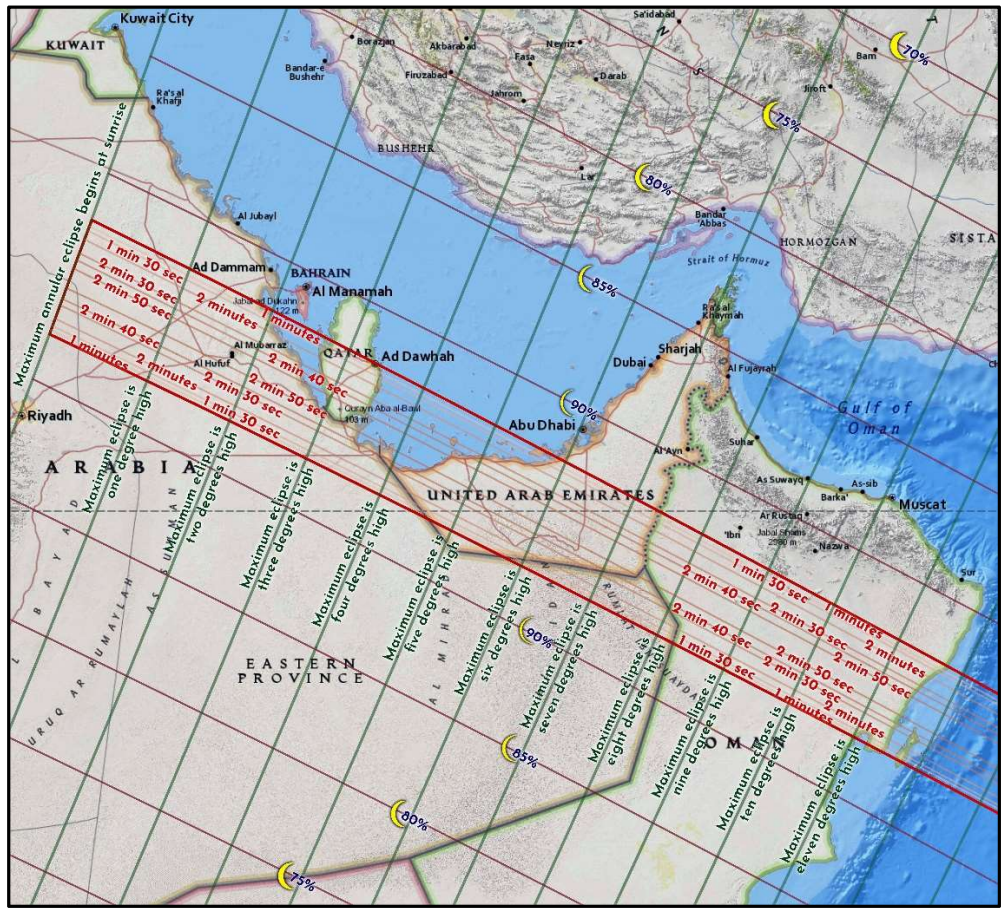

**Figure. 3**. Map showing the penumbral track of the annular solar eclipse of 26 December 2019, through Saudi Arabia, Qatar, United Arab Emirates and Oman (map by Michael Zeiler; reproduced with permission).

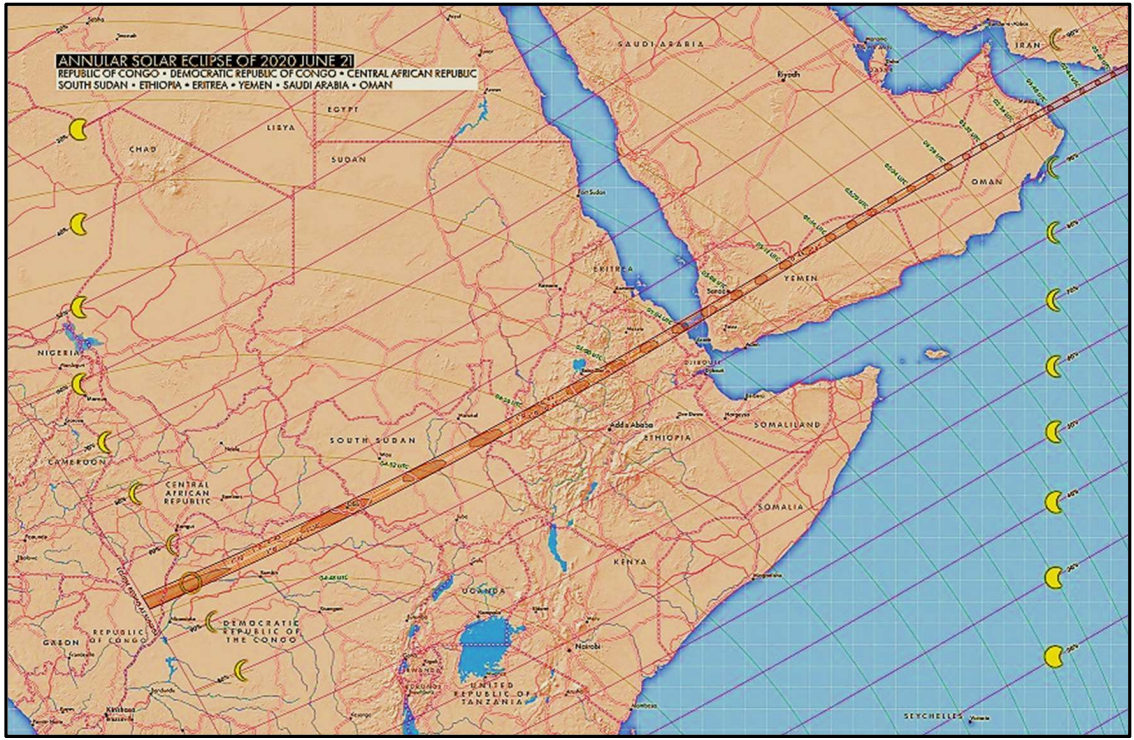

**Figure 4**. Map showing the penumbral track of the annular solar eclipse of 21 June 2020, through Republic of Congo, Democratic Republic of Congo, Central African Republic, South Sudan, Ethiopia, Eritrea, Yemen, Saudi Arabia and Oman. [map by Michael Zeiler, greatamericaneclipse.com (reproduced with permission)].

### 3.2. Geographical

Riyadh (24.86° N; 46.40° E, 764 m asl.), lying in the central region of the Arabian Peninsula, is the capital of the Kingdom of Saudi Arabia and is considered one of the most polluted cities of the Kingdom (Al-Rajhi *et al*., 1996; Alharbi *et al*., 2014, 2015).  In winter air pollution is related to falling dust, which is a frequently phenomenon between January and March (Modaihsh & Mahjou, 2013).  Riyadh is also characterized by low humidity and large seasonal temperature differences, particularly in the summer; the temperature difference is affected by the arid conditions due to proximity to the Empty-Quarter Desert. Al-Hofuf city is the major urban region in the Al-Ahsa oasis in the eastern province of Saudi Arabia, located on the center line of the 26[th] December eclipse path. The city is characterized by high humidity, especially during the month of December (an average of 56%).

Al-Hofuf eclipse was close to the Four-mountains site (25° 17′ N, and 49° 42′ E), which is located in a remote area of Al-Hofuf region (Figure 5). While there was some haze on the horizon, the weather conditions were optimal to see the whole eclipse. For Riyadh, the observations were made at the King Saud University observatory, while the weather parameters measurements were recorded at King Abdulaziz City for Science and Technology (KACST) campus (24° 43′ N, 46° 40′ E; altitude: 613 m).

A description of soil texture and topography of the Arabian Peninsula area where the 26 December 2019 was viewed (Al-Hofuf) is given by Nelli *et al*. (2020); their figure 1 shows that around Four Mountains Camp in the upper side of the eclipse´s central line the soil is made of sandy loam and loamy sand; and in the lower or opposite side of this line the soil is made of sandy clay and clay loam.

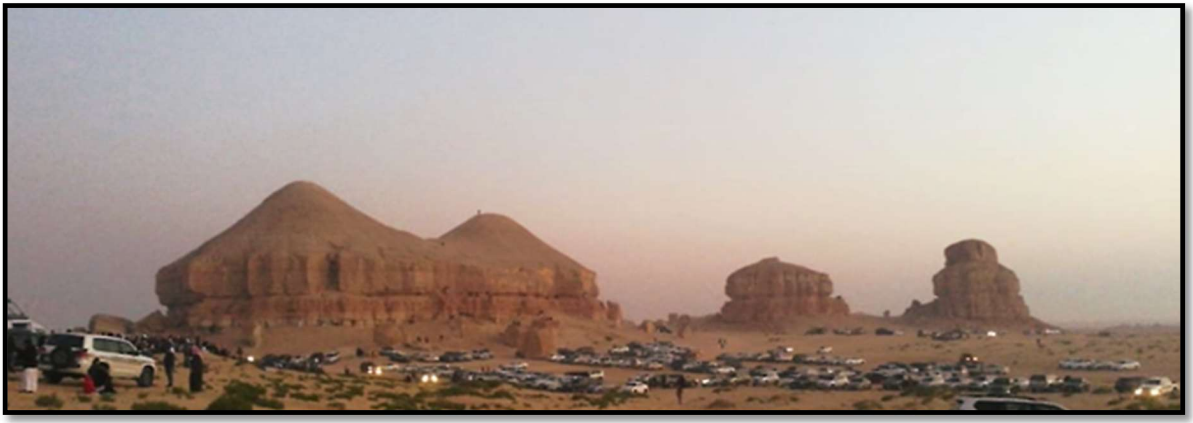

**Figure. 5.** A panoramic view of the Four-mountains site at Al-Hofuf City where observations were made for the eclipse of 26 December 2019, just before sunrise.

### 3.3. Climatological and Meteorological

As described by Jay Anderson in his website https://eclipsophile.com/annular-solar-eclipses/ase2019/, Arabian Peninsula climatology indicates that for December winter produces in this area colder weather and occasional storms generated by a northwesterly wind (known as "Shamal"), which follows the passage of a dry cold front causing frequently and significant dust storm (Modaihsh & Mahjou, 2013). Caused by northeasterly winds that blow across the Persian Gulf from Iran, or due to the higher terrain in Saudi Arabia and also in Iran, this wind is directed down the Persian Gulf. However, these winter dust storms are not common excluding the coast of the peninsula where it may occur 2 – 3 times per month and last 24-36 hours. Shamals lasting longer than this (3-to-5 day) are still less with a frequency of one or two events per winter season, being the U.A.E. – Oman border the region most affected. A Shamal can transport into the air dust in a range of many kilometers producing some haziness of the atmosphere owing to suspended particulates, which can affect the surface visibility.

Brought on by occasional weak disturbances from Africa and the Mediterranean that reach the area in winter, cloud levels are moderately high in this region. Average December cloud amount along the eclipse track, derived from 15 years of Aqua & Terra satellites observation, shows a rapidly improvement of cloud climatology within the southeastward-trending eclipse track. According to measurements made by these satellites cloudiness drops to the lowest level along the track just 7% crossing Oman. From the observed levels of cloudiness, the forecast for the December eclipse day looked very good but not as optimistic due to some influence of the Arabian Sea, which produces cloudier skies than in land coastal cities than inland deserts. According to the data, minimum cloudiness occurs in the stony desert of Oman's interior characterized by a general lack of vegetation giving the sunny and waterless character of the climate in that zone.

Regarding the atmospheric condition at meso-scale level on 26 December 2019, in the region of the Arabian Peninsula, it was estimated at 06:00 (local time) by Nelli *et al*. (2020) by applying ERA-5 reanalysis data at 0.25° x 0.25° spatial resolution. Their results are presented in Figure 6, which gives the 10-m horizontal winds, surface temperature, total cloud cover, and precipitable water (i.e., total amount of water vapor in the atmosphere) just before the local sunrise. Their analysis showed that, except for the coastal parts of Oman, Qatar, and the neighboring Saudi Arabia, skies were generally clear in the region (center panel of Figure 6), which coincides with the December climatology described above. Low atmospheric moisture (right panel of Figure 6) induced strong radiative cooling producing nightime temperature values below 10 °C in some areas as shown in the left panel of Figure 6. A wind inspection indicates that northeasterly winds prevailed over the Arabian Sea, which penetrated inland Oman and eastern Saudi Arabia.

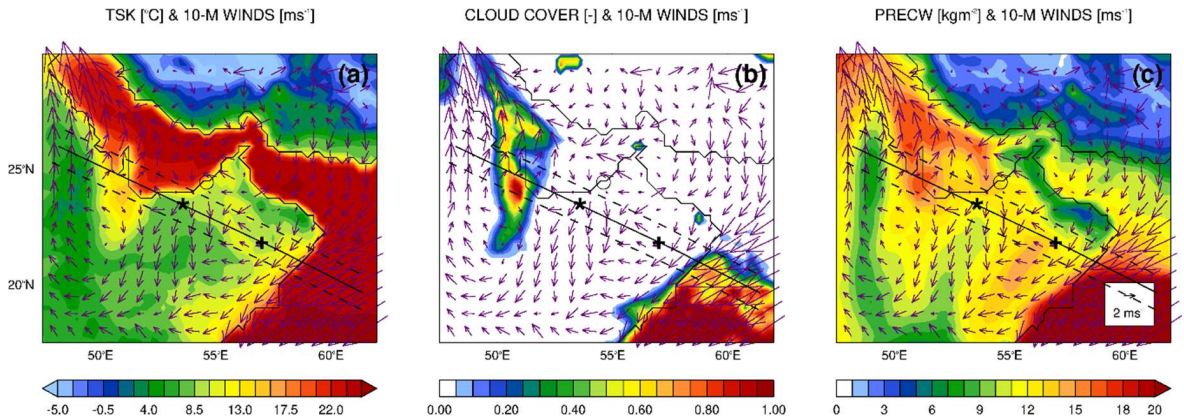

**Figure. 6.** Horizontal wind at 10-m height [vectors (m-s⁻¹)] along with (a) surface "skin" temperature [TSK, (°C)], (b) non-dimensional total cloud cover, and (c) precipitable water [PRECW (kg-m⁻²)], from ERA-5 reanalysis at 06:00 (local time) by Nelli *et al*. (2020), on 26 December 2019, in part of the Arabian Peninsula where the 2019 ASE was observed. Black star (in the U.A.E.) and black plus sign (in Oman) indicates eclipse observation sites (stations) used by Nelli *et al*. (2020) (reproduced with permission).

The forecast for the 2020 June eclipse was given by Jay Anderson in his website https://eclipsophile.com/ase-2020/. Except for a part of Yemen, most of the eclipse track crossed the Arabian Peninsula (Figure 4). That region is characterized by extremely low average cloud amounts of between 2% and 5 %. Satellite-based measurements reveal that June's average cloudiness at 10:30 is greatest over Yemen's Western Plateau, where the terrain rises as high as 3000 m. In Yemen, summer months come under the influence of the southwest monsoon winds, bringing humidity and small amounts of rain to the highlands and a general cloudiness to the rest of the country. Most of this cloud is convective in nature and builds in the afternoon, leaving the mornings and the eclipse with a generous, sunny climatology. Average morning cloud peaks at just over 20 %, about half of that measured in the early afternoon. Inland, 250 km from the Red Sea coast, at Ma'rib, the eclipse track leaved the Western Plateau and moved out onto the Rub 'al Khali in Saudi Arabia, which is a landscape of sand dunes receiving cold fronts with strong northwest winds bring the Shamal; this is more often associated with Iraq and the Persian Gulf countries. The few ground-based measurements in this region show that the percent of possible sunshine ranges from 68% to nearly 80% along the track, with the lowest values over Yemen and the highest over Oman. Measurements reveals that daytime highs average between 40°C and 43°C across the Peninsula, except in Yemen, where afternoon cloud cloaks the sun. Record highs reach almost 50°C. And though humidity is low in the desert, coastal regions of both Yemen and Oman are noted for their sultry weather when winds blow onshore. Using average temperature and RH values at eclipse time gives humidex scores of 38°C Muscat in Oman and 40°C at Al Hudaydah in Yemen; maximum daily values are considerably higher. Hot and dry desert conditions would persist until the track encounters the Northern Oman Mountains, whose peaks reach as high as 3000 m. These is a small increase in cloudiness over the mountains, but here again, it is primarily an afternoon phenomenon.

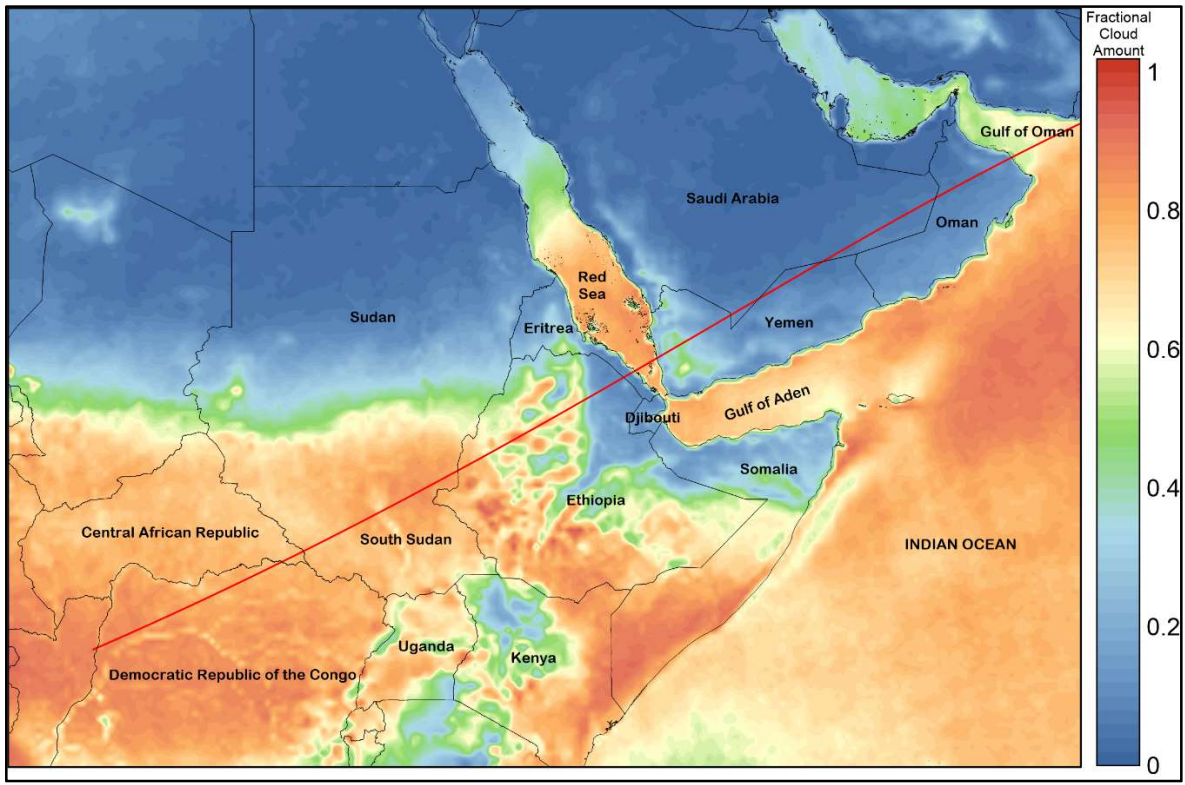

**Figure. 7**. Map of average 10:30 (local time) June cloud cover along the eclipse track over Arabian Peninsula. Map is based on 18 years of satellite observations from the Terra satellite [https://eclipsophile.com/ase-2020/] (reproduced with permission).

It is important to note that the weather conditions were optimal for both eclipse. Clear skies and cloudless prevailed during the course of observations (even during the whole days). In both sites the weather is currently characterized by very hot, dry, and long summers and cold and dry winters. The sky is mostly clear along the year at both sites.

Al Hofuf, with a Köppen climate classification of BWh, has a hot desert climate with long, very hot summers and mild, short winters. In December, between 1985 and 2010, some of its climate data is presented in Table II as well as for Riyadh in the same interval but in June. Riyadh City has also a hot desert Köppen climate classification with long, extremely hot summers and short, very mild winters. The average high temperature in August is 43.6 °C. The city experiences very little precipitation, especially in summer, but receives a fair amount of rain in March and April. It is also known to have dust storm during which the dust can be so thick that visibility is under 10 m.

**Table II**. Some climatological pertinent data of Al-Hofuf (Al-Ahsa) and Riyadh between 1985 and 2010

| | Al-Hofuf (Al-Ahsa) | Riyadh |
|---|---|---|
| Station name / Number | Al-Ahsa / 40420 | Riyadh Old / 40438 |
| Month | December | June |
| Record high (°C) | 32.5 | 47.2 |
| Average high (°C) | 23.4 | 42.5 |
| Daily mean (°C) | 16.6 | 35.7 |
| Average low (°C) | 10.5 | 28.0 |
| Record low (°C) | 0.8 | 21.1 |
| Average rainfall (mm) | 21.1 | 0.0 |
| Average RH (%) | 56 | 11 |
| Average rainy days | - | 0 |
| Mean monthly sunshine | - | 328.2 |
| Percent possible sunshine | - | 80 |

Note: Data taken from the surface annual climatological reports issue by the Environment Protection, National Meteorology & Environment Center of the Presidency of Meteorology & Environment Protection, Ministry of Defense & Aviation, Kingdom of Saudi Arabia.

## 4. Equipment and data

At both sites, air temperature was obtained with a sensor attached to another sensor, which measured solar radiation; both are described in details in Maghrabi *et al.* (2009). The data were acquired with a good time cadence of 10 s for the Al-Hofuf eclipse while for the Riyadh event, using a XR5-8-A-SE data logger manufactured by Pace Scientific, this cadence was every 1 minute for eclipse day and 10 min for the pre-eclipse day; In Al-Hofuf the sensor and the logger were placed on a horizontal surface in open air away from any obstruction or obstacles. At Riyadh City the logger and the shielded sensor were placed in open air, directly exposed to sunlight, at the roof of the radiation detector lab located at King Abdulaziz City for Science and Technology (KACST) campus (6 m above ground level; see Figure 8). The logger has internal sensors to capture humidity and air temperature data. The accuracy of the measurements with the logger's sensor is ± 2% for RH and 0.15 °C for temperature at 25 °C. The temperatures from both the sky radiometer and the logger were very close to each other with mean difference less than 0.1 °C. In the present study, air temperature and humidity data were taken from the logger.

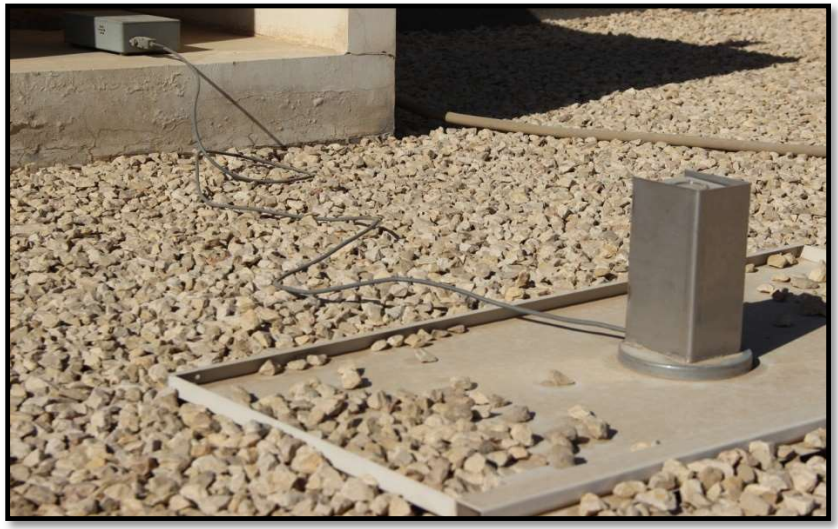

**Figure. 8.** The sky radiometer shown in this image (bottom right) was installed at KACST campus 6 m above the ground level, from where the observations of air temperature and RH in open air were made during the eclipse day of 21 June 2020. This radiometer has inside a shielded temperature sensor that is connected to the logger. Note the artificial nature of the surface where this equipment was and the environment surrounded it, which was not the case for the eclipse of 26 December 2019, in which the equipment was installed on a natural desert environment in open air.

## 5. Results and analysis

### 5.1. Al-Hofuf

Firstly, in Figure 9 we report results, for the Al-Hofuf annular solar eclipse, illustrating the temporal evolution of the temperature [Figure 9 (a)] and RH [Figure 9 (b)] in the time interval of 04:00-10:00.  In order to detect any trend variation due to the eclipse, the data are compared with non-eclipse data, which in this case was that of the day before eclipse on 25 December 2019 (blue short-dashed lines in Figure 9).

For both variables during the eclipse and non-eclipse days in this interval the curves display an almost stable evolution in magnitudes until reaching the eclipse's annular phase. Owing to the occurrence of sunrise prior and close to annularity (in the middle of the eclipse phase), the effects of the rising Sun and the eclipse combine to alter the progression of the curves.

On close inspection of these changes with respect to a non-eclipse day, we observe how the difference in our micrometeorological variables evolve in time, which are shown in lower panels of Figure 9. From lower panel (a) of this figure the temperature difference, defined as $\Delta T_{26-25} = T_{26} - T_{25}$, appears to retain a constant value of approximately $2^{\circ}C$ until crossing the annular phase where this difference clearly drops by an amount of 1 °C, increasing gradually up to the final phase of the eclipse. This change is basically due to a delay in atmospheric temperature response, which increases after sunrise compared to that of the non-eclipse day, and that affects the temperature curve slope producing this behavior. A similar response is observed in RH difference, defined as $\Delta RH_{26-25} = RH_{26} - RH_{25}$, in the lower part of graph of Figure 9 (b), with a rather visible well-recognized increase trend up to the end of the eclipse of almost 3%.

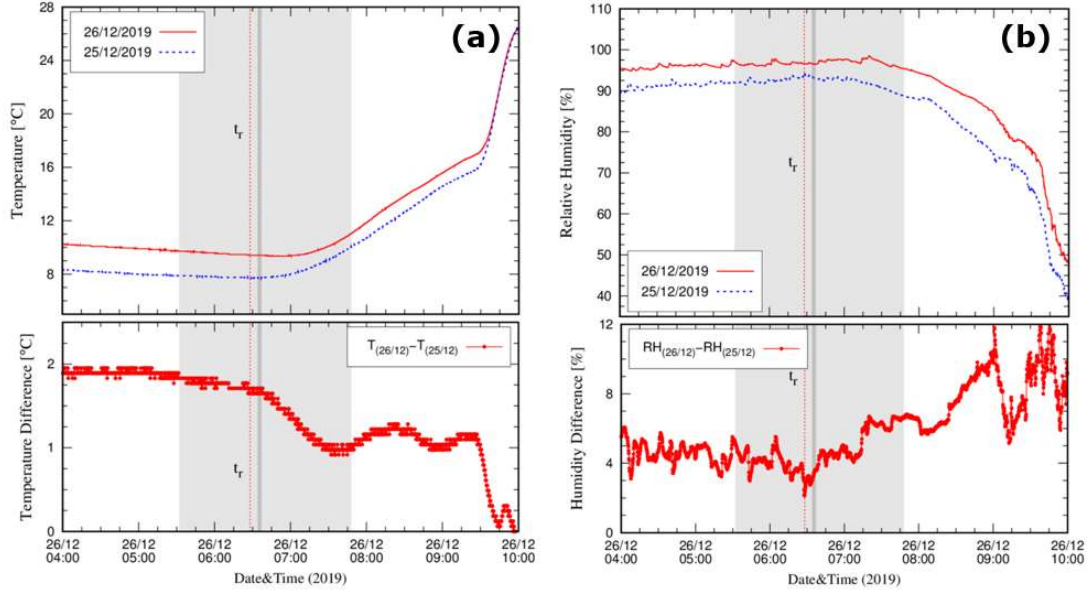

**Figure 9**. Temporal evolution of the temperature (a) and RH (b) at Al-Hofuf between 04:00 and 10:00 on the annular eclipse day (red continuous line), 26 December 2019, compared to previous non-eclipse of 25 December 2019 (blue short-dashed line). The vertical dotted line indicates the sunrise time, $t_r$. The shaded area refers to the solar eclipse phase, while the shaded area between vertical lines indicates the short phase of the annularity. Note the high magnitude values of RH at the Al-Hofuf site during this season, which is over the average shown in Table II. The bottom panels highlight the differences between eclipse and non-eclipse days, $\Delta T_{26-25}$ and $\Delta RH_{26-25}$.

In addition to the above results, we estimate VP variation during the eclipse in Figure 10 (a), which has been obtained using data of temperature and RH by applying the relation,

$$e = RH\ e_s(T) = RH\ a\ exp\ [bT/(T + c)]\ ,\qquad(1)$$

where $e$ is actual vapor pressure and $e_s(T)$ is the saturation vapor pressure at actual temperature T (in Celsius-degree) (Idso & Jackson, 1969; Buck, 1981; Bohren & Albrecht, 1998; Campbell & Norman, 1998), with $a$ = 0.611 (kPa), $b$ = 17.5 and $c$ = 241 °C. The above relation stems obviously from considering the vapor pressure deficit as $e_s(T) - e = e_s(T)$ (1 – RH). Apparently, there is no signal of influences or perturbations on VP during this eclipse at Al-Hofuf. The VP remained approximately constant over the most part of the eclipse period until 07:13 when it began to increase steadily for the rest of the measurements. Although there was a small increase in VP during the latter part of the eclipse measurement period, this was most likely not related to the eclipse. Although a statistical analysis of the typical variance in VP inferred from numerous observations would be necessary to confidently determine that these changes were insignificant, a comparison with the day before supports this finding.  A similar non-effect was found by Pleijel (2008) observing the late-morning TSE of 29 March 2006 at Side, Turkey, he explains that some time eclipse effects on air humidity are reported to a much lesser extent than effects on temperature than was found by Kolarž *et al.* (2005) during the solar eclipse of 11 August 1999, in Belgrade, Yugoslavia, and Ahrens *et al*. (2001) observing earlier the same eclipse but in southwest Germany. In passing, Huzimara (1949) observing the mid-day solar eclipse of 9 May 1948 from Mt. Huzi in Japan, at 3780 m asl., did not detected any change in VP.

Under a different perspective this non-effect on VP is observed in Figure 10 (b), where we have plotted the difference, defined as $\Delta VP_{26-25}= VP_{26} - VP_{25}$, during the same period of time as in Figure 10 (a). Between sunrise (at 06:28:07) and last contact (at 07:47:57) there is no appreciable effect of the eclipse over the trend of this difference, indicating that mechanism described in Figure 1 did not take place or was minimum as to be influential at this time of the morning. From upper panels (a) and (b) of Figure 9, one can see that the impact of the eclipse on temperature and RH, respectively, was negligible and so over VP [see Eq. (1)].

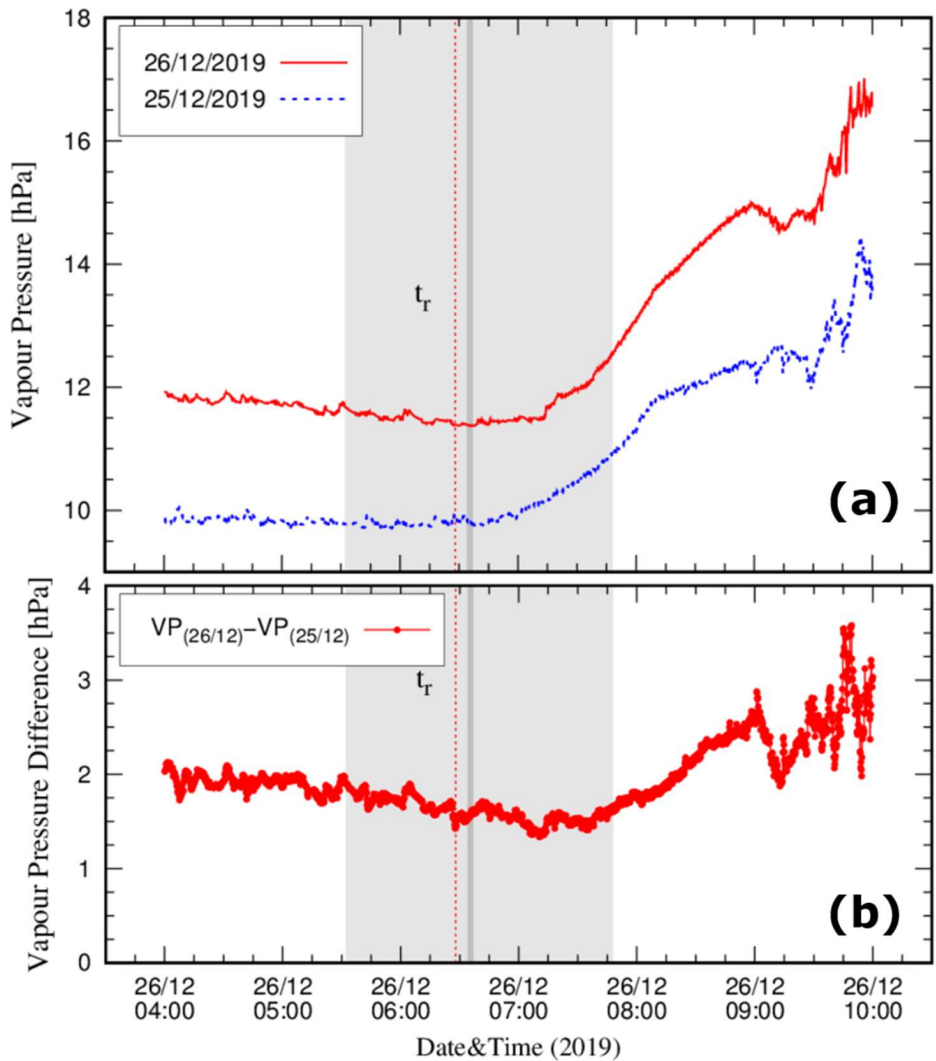

**Figure. 10.** Upper panel (a). Vapor pressure variation at Al-Hofuf on 25 and 26 December 2019 between 04:00 and 10:00. Both patterns are similar except that eclipse day values of VP were greater than those of the non-eclipse day. No signal of any eclipse effect was detected during the observation

that took place at that site between 06:28:07 (at sunrise) and 07:47:57. Lower panel (b). Difference in VP at Al-Hofuf between 26 December 2019 (eclipse day) and the day before, from 04:00:57 to 09:59:57. This variable tended to decrease as long as sunrise was approaching; it continued in this way but under the presence of an annular eclipse in progress at dawn. At some point, such difference began to increase as shown from 07:40 to past 08:52 when strong oscillations appeared. There was no any perturbation or effect produced by this eclipse over VP.

Another view, to have an idea of the non-impact effect over VP by this eclipse, is to examine the VP anomaly and compare it between eclipse day and a non-eclipse day (the day before in this case). The anomaly is defined as the average of VP values, calculated for determined time interval, minus the value as a function of time in that interval. From Figure 11 we can observe that the anomaly for both days, between sunrise and last contact, is approximately the same, with a first part with positives values and a second part with negative values; however, a small discrepancy is noted in the eclipse curve, between 7:01 and 7:15, which is not related to the eclipse.

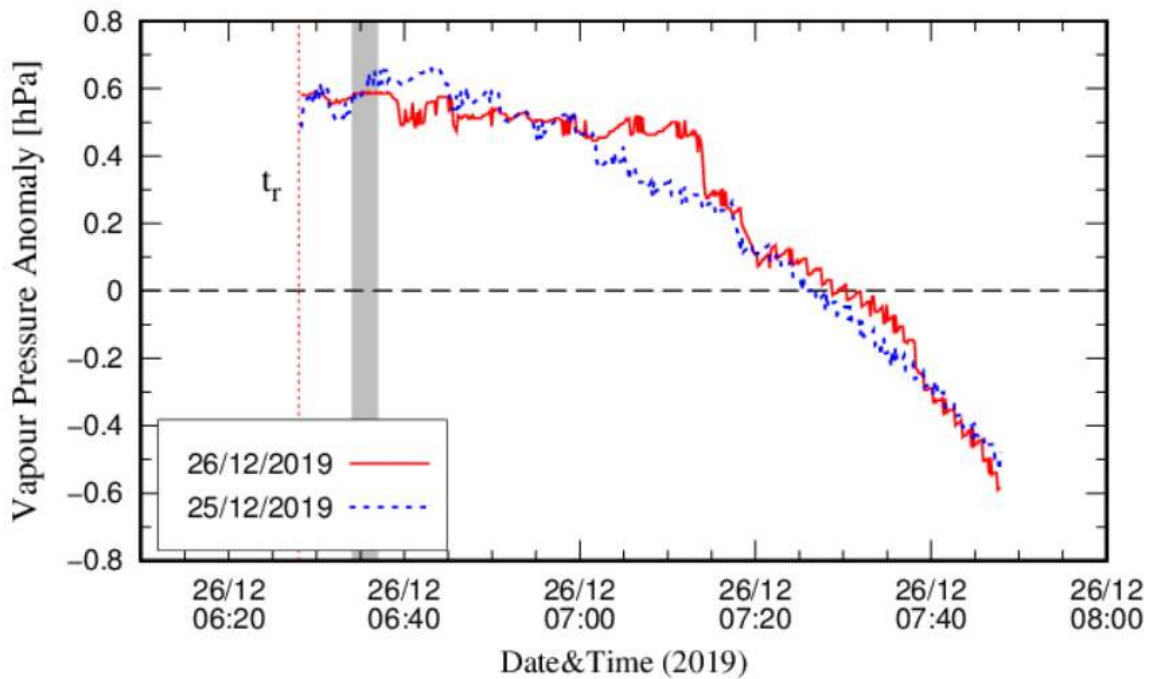

**Figure. 11**. Anomalies of VP for the eclipse day and the day before at Al-Hofuf site. A comparison between both curves suggests a non-effect of the eclipse over VP during this event from sunrise to last contact in accordance with results shown in Figure 10 (b).

Al-Hofuf, from Table II, had a 53% RH average in December between 1985 and 2010. For the same month and interval this site had an average of 21.1 mm in rainfall both averages indicating a typical dry weather therein. With the surface so dry, the vertical gradient in the water vapor was likely weak; therefore, even if subsidence via the mechanism described in Figure 1 were to have occurred in these annular eclipse circumstances, changes would likely have been beyond our detection. Unfortunately, no atmospheric pressure and wind measurements were made to investigate this further.

*5.2. Riyadh*

As for the Riyadh eclipse, with its 73% occultation, the circumstances and associated effects are different from those reported for the Al-Hofuf annular eclipse. Although this eclipse was also a morning event, however, there was sufficient time between sunrise and first contact of about 2 hr 05 min (see Table I for more details) for temperature begins to increase. Figure 12 depicts the temporal variation of temperature [panel (a)], and RH [panel (b)], between 02:00 and14:00. The temperature depression is highly noticeable over the eclipse phase [shaded area in Figure 12 (a)]. We note that the temperature continued to increase for about 18 min after the first contact, most likely as a response of the pre-existing diurnal temperature increase (Clark, 2016). To detect the impact of the solar eclipse on the temperature and RH, we compared the data from 02:00 to14:00 with those of a non-eclipse day. Results are highlighted also in Figure 12 (a) – (b), where our measurements of post-eclipse day (i.e., 26 June 2020) are plotted in blue short-dashed lines. The above discussed effects appear clearly to be absent for an "usual" day, evidencing a variety of possible irregularities in meteorological variables that might be attributable to the eclipse and related effects.

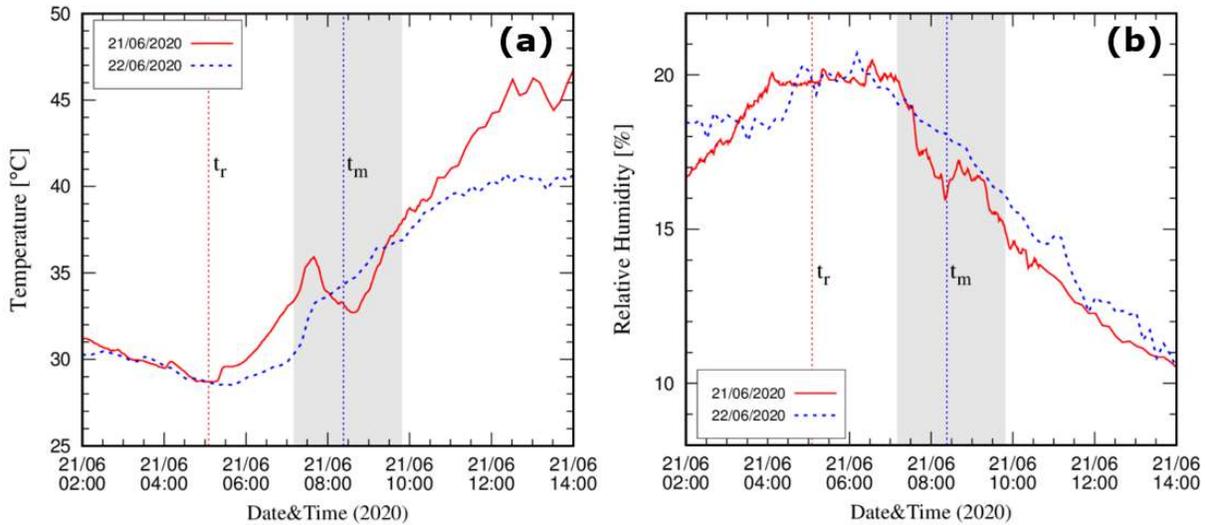

**Figure. 12**. Temperature (a) and RH (b) zoomed-in views between 02:00 and 14:00 during the Riyadh partial eclipse of 21 June 2020. Sunrise and maximum time are shown as vertical dotted lines. Shaded area refers to the eclipse phase. Panel (a) highlights some characteristics related to the observed temperature drop (see text for more details). These results (red continuous lines) can be compared with a non-eclipse day of 22 June 2020 (blue short-dashed lines).

A black short-dashed line to simulate a hypothetical non-eclipse temperature trend is shown in Figure 13 (a). This gives a characteristic dip temperature of about 4°C over a shape of full width at half-maximum (FWHM) of ~ 65 min, whereas a value of about 3.2°C is estimated as a maximum-to-minimum fall in an interval of almost 56 min. The first dip temperature is known as temperature absolute anomaly, and the second dip as temperature linear anomaly (Clark, 2016; Peñaloza-Murillo *et al*., 2020). Additionally, a time lag (time between maximum obscuration and the reached minimum) is distinguishable; we estimate to be it about 15 min.

The progression of the RH is shown in Figure 13 (b). Note the low magnitude values of RH characterizing Riyadh City for this month of the year (see Table II). Starting at sunrise we remark the presence of perturbations in the recorded values, while a steep decline coincided with the beginning of the eclipse phase governed by a total fall of about 5% over the eclipse time interval (i.e., 1CJ to 4CJ). An increasing trend right after maximum is noticeable; most likely it was caused by decreasing air temperature owing to a gradual shading of the Sun by the Moon. At last contact 4(CJ), the RH curve returns to display a smoothie trend.

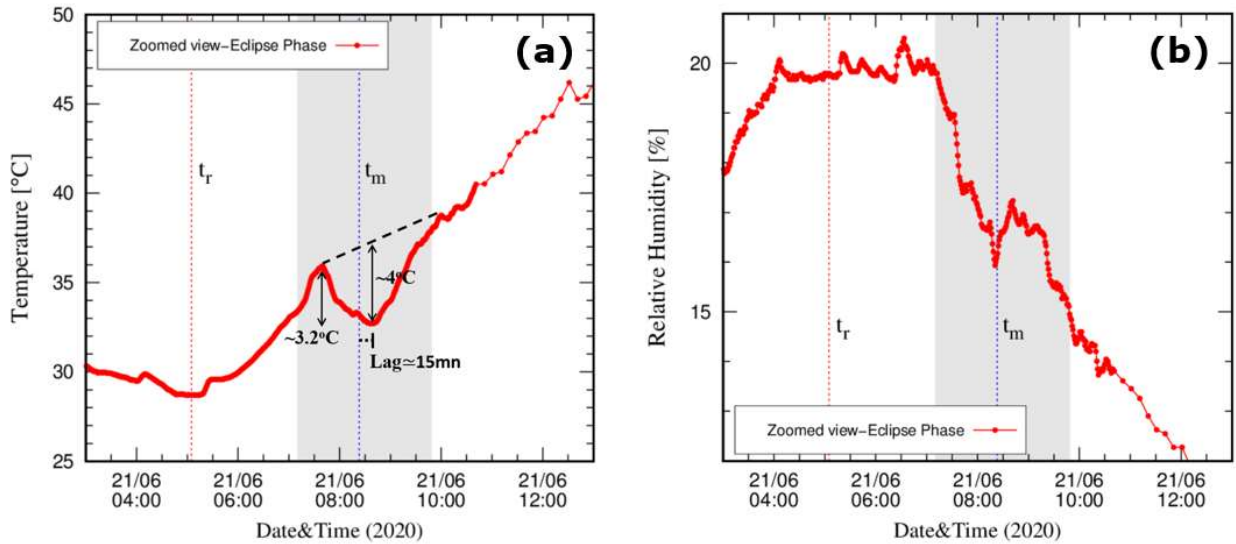

**Figure. 13**. Temperature (a) and RH (b) zoomed-in views, in the 02:00 am to 14:00 am time interval during the Riyadh partial eclipse of 21 June 2020. The sunrise occurrence and maximum time are shown as vertical dotted lines. Shaded area refers to the eclipse phase. Panel (a) highlights some characteristics related to the observed temperature drop (see text for more details).

By applying Eq. (1) we present in Figure 14 results of VP change. We note clearly the impact of this partial eclipse on this variable on 21 June 2020, at Riyadh. For comparison purposes the curve for the day after is included [as in Figure 12 (b) for RH] as well as that for the day before. To understand the response of VP to the eclipse we have to turn our attention to the response of RH for both dates [Figure 12 (b)]. For the normal day after the eclipse (June 22), during the same period of time corresponding to the eclipse time (June 21), say, between 07:10 and 09:49, RH decreased steadily; during the partial eclipse this change occurred until 08:20:36 (to a value of 15.94 %). From this point onwards, as a response to the eclipse, RH began to increase until 08:41:36 (to a value of 17.24 %) when it began to decline again quite similar to the day after. Therefore, the eclipse effect on RH was noticeable from 08:20.5 to 08:41.5, approximately.

As a consequence of the above change, it is observed that during the eclipse VP decreased between 07:35 and 08:20 product of a falling of temperature and RH as well. Next, from 08:21:36 this variable began to increase as it is expected during an event like this until it reached similar values of the day after past last contact. For the day after, VP had an increasing behavior in contrast with the seesaw character seen on VP detected during the partial eclipse. The latter also apply for the day before on 20 June 2020.

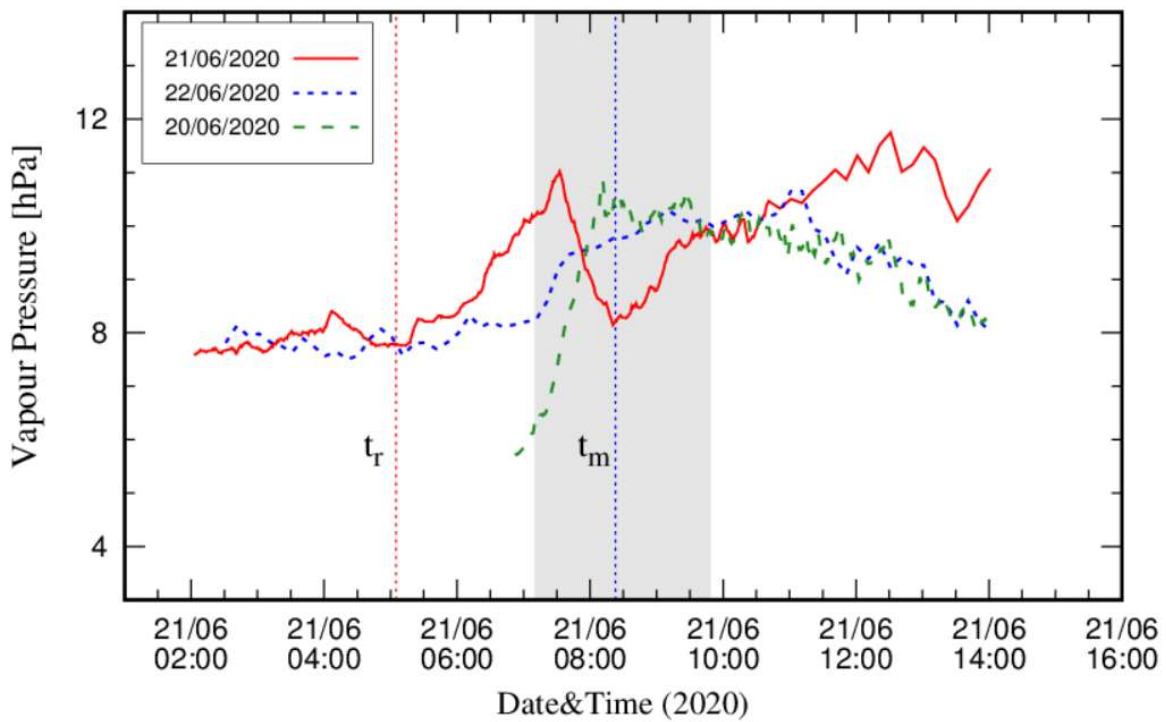

**Figure. 14**. In this comparison of VP at Riyadh during the partial solar eclipse on 21 June 2020 (07:10 – 09:49) and the day after (22 June 2020) (blue short dashed-line) at the same interval, the effect of the eclipse is seen (red solid-line). From a certain point (at 08:20.5) onwards it produced an expected increase of VP until 08:41.5, very close to the final phase of the eclipse. A third curve (in green dashed-line), corresponding to day before the eclipse (June 20, from ~ 06:50 to 14:00), has been included to confirm such an effect.

## 5.3. Eclipse related-dynamical variations and inter-comparison

Aside from the significant differences in magnitudes and general trend of the meteorological data for the two eclipses, it is important to evaluate closely the related-dynamical variations via rate calculations. The computation of rates implies estimating first derivatives of the studied variables with a "moving-average" smoothing technique that has been applied for a better curve-evolution visualization. Our results are depicted in Figure 15 (a) – (b) for the Al-Hofuf eclipse and Figure 15 (c) – (d) for the Riyadh eclipse. Furthermore, we calculated the rates for non-eclipse days (blue short-dashed curves) for comparison. Horizontal dotted lines indicate zero-rate level to help in the interpretation of the variations.

The two eclipses related rate-curves are evidently different in various aspects. For the December's eclipse, because sunrise occurred during the eclipse, before the annular phase, the effects are rather insignificant compared to June's eclipse. On the one hand, we expect that after sunrise the temperature increases, which can be seen for both curves (eclipse and non-eclipse days) with increasing positive rates [Figure 15 (a)]. The effect of the eclipse then is imprinted as a decrease of the observed trend compared to non-eclipse day, which is distinguishable right at the annularity and later-on as a distinct difference between the two curves (dashed and continuous lines). An average difference rate of about 0.022 °C/min (~1.32 °C/hr) is actually measured between the two curves, being the non-eclipse curve higher than the other, lasting almost until the end of the eclipse. Outside the eclipse phase the two curves/rates are very similar in both trend and magnitude.

In addition to what it is noted in temperature, RH rates (RHR) were very similar (although subjected to more fluctuations) for both days. After sunrise, we envisage that RHR has negative values (i.e., decreasing), which in fact we observed in our rate curves. The only difference is that due to the eclipse effects the humidity for the eclipse day should decrease slowly compared to the non-eclipse day, which we can note in our rate curves [overall larger values in the continuous-curve compared to dashed one up to the end of the eclipse phase [Figure 15 (d)].

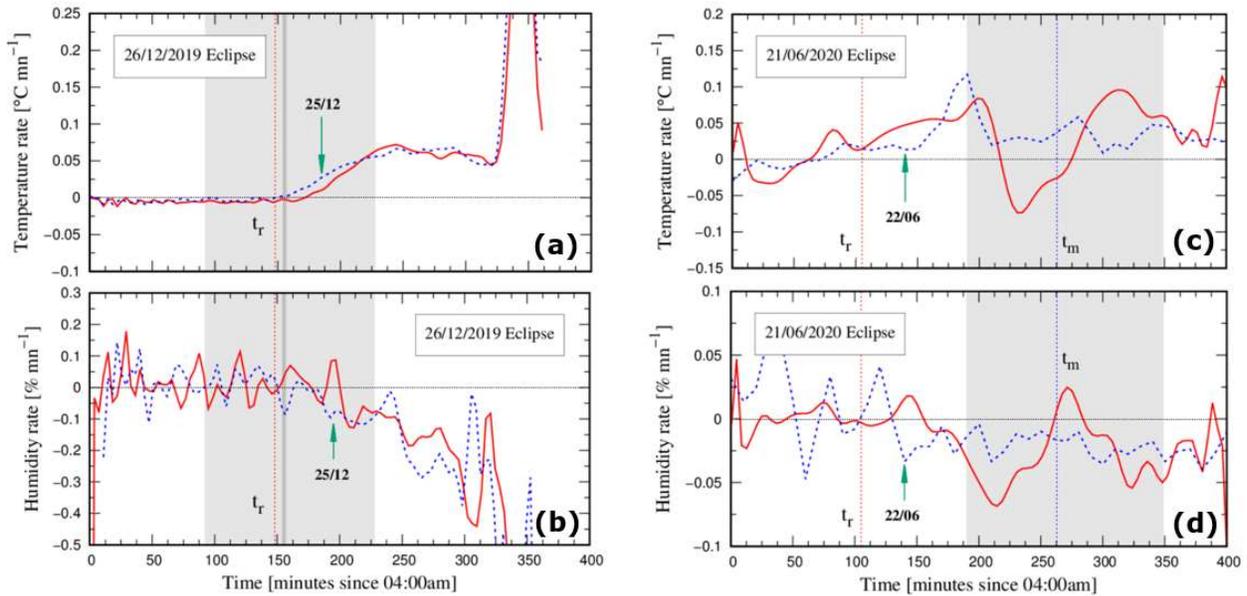

**Figure. 15**. Temporal evolution of the computed rates of temperature and RH for Al-Hofuf eclipse [panels (a) and (b)] and for Riyadh eclipse [panels (c) and (d)]. For both eclipses the computed rates for non-eclipse days are also reported (blue short-dashed blue curves). Horizontal dotted lines indicate zero-rate level. The curves have been smoothed using "moving-average" technique for a better curve evolution visualization.

## 6. Inter-comparison among studies of other eclipses in desert or similar environments

In this section we will try to compare the impact of the eclipse of 26 December 2019, at Al-Hofuf, and of 21 June 2020, at Riyadh, to other solar eclipse impacts observed in desert and similar environments reported by other authors. In particular, the impact of the eclipse of 26 December 2019, at Al-Hofuf will be considered separately for a comparison with those observations published by Nelli *et al.* (2020) on this eclipse.

### 6.1. Previous studies

Tables III contains useful information of temperature, RH and VP response to solar eclipses (total, annular and partial) in different desert or semi-arid environments located in southern U.S.A., northern Africa and in the Near East, within a region comprised between (20° 26'- 36° 22') N and (106° 20' -  40° 00') W-E (see map of Figure 16), to be inter-compared. This information has directly been taken or estimated from the works of Jaubert (1906), Klein & Robinson (1955), Anderson & Keefer (1975), Eaton *et al.* (1997), Hassan *et al.* (1999), and Hassan & Rahoma (2010).

The eclipses of Table III have orderly been arranged according to the climatological season sequence. Thus, the first eclipse is that occurred as a partial in Israel in the second half of the morning of 25 February 1952, which was observed from a network of meteorological stations (Klein & Robinson, 1955). In particular, air temperature was measured only at Tel Aviv (Hakirya) and Eilath, a site in the southernmost desert region of the Araba Valley (on the shore of the Gulf of Aqaba), where RH and VP were also measured. With an occultation of 75.30% at Eilath, the impact on these three variables were evident but small. A decrease of ~1.0 °C was observed, in the range of 18.0 °C – 17.5 °C, along with an increase in RH past the maximum until fourth contact breaking, in this way, the usual or normal morning decreasing trend as long as the temperature increases. As for VP it increased between first contact until middle or maximum when, from that point, began to decrease accordingly to the previous and subsequent days also measured. In Tel Aviv, with an occultation of 69.19%, the impact on temperature was not so remarkable (less than 1 °C) such that it can be considered negligible. The approximated values presented in Table 3 for these sites were extracted graphically from the plots published by these authors in their booklet. By applying Eq. (1) to temperature and RH approximate values the results are quite close [10.4 (1C), 9.27 (Max), 10.80 (4C)] hPa.

Next, is the 29 March mid-day almost total eclipse (99.98%) of 2006 at Tubruq, Lybia, (third line in Table III) near the Mediterranean coast with a background considered to be a desert (Hassan & Rahoma, 2010). The impact on temperature, RH and VP were also evident but not so large. These authors found a small dip of 1.5 °C in temperature along with a RH increase from 62% (at first contact) to 82% (at maximum eclipse), and from that value it decreased to a value of 78%; therefore, RH varied as expected. Vapor pressure, calculated via Eq. (1), showed an increased from 14.04 hPa (at 1C) to 19.38 hPa (at 4C). In comparing the Israeli eclipse (at Eilath), dry and a little cold, with the Libyan eclipse (at Tobruq), mild and humid, both close to the sea, we observe the effect mostly due to a change of RH and VP.

Crossing the Atlantic, on 10 May 1994, an annular solar eclipse was observed in a desert site of the Tularosa Basin of New Mexico, U.S.A., characterized by low brush (predominantly mesquite), associated desert grasses, and other herbaceous plants indigenous to the Chihuahuan desert. It was a spring morning eclipse with an arid continental general climate with annual precipitation of 175 to 275 mm per year. With an occultation of 86.65% at annularity Eaton *et al*. (1997) reported only air temperature measurements establishing that the data showed a stable near-surface air from the surface to 20 m above the ground level (the site has an altitude of 1200 m asl), from about 45 min centered around the time of maximum eclipse. Thus, the impact of the eclipse at that height was unnoticed (see their figure 3), but at the 4 m level air temperature decreased about 3 °C during the eclipse. No information was given on RH and VR. Considering the season of the year the dip in temperature contrasts with those in Israel and Lybia.

Back to Africa, Chinguetti, Mauritania, was the site chosen by Anderson & Keefer (1975) to observe the longest total solar eclipse of the twentieth-century, with a totality duration at that location of 6 min and 17 s, on 30 June 1973, at that site [totality duration went up to 7 min 4 s in Niger, though the site was not useful for eclipse observations because of its location a hundred kilometers north of Timbuktu, in spite of the reconnoitering of the site three years earlier by one of us (JMP)]. During the time around totality in Mauritania, a dust storm came up, diminishing transparency to about 10%. The Chinguetti has a desert climate with virtually no rainfall during the year. The Köppen-Geiger climate classification of this site is BWh. In June the average high temperature is 40.5 °C with an average precipitation of 2 mm. These authors reported absolute anomalies ($T_{max}$-$T_{min}$) of 3.5 °C at 0.3 m and 2.5 °C at 6.75 m and 13.5 m above the surface. The values at 0.3 m and 6.75 of 3.5 °C and 2.5 °C, respectively, seems to coincide to the value of 3 °C given for the May New Mexico desert eclipse but at 4 m. No data for RH and VP were given for this eclipse at Chinguetti.

Moving on to northern Africa, a total solar eclipse was observed at the beginning of the twentieth-century in Constantine, Algeria, by Jaubert (1906) on 30 August 1905. Constantine has a Mediterranean climate (Köppen climate classification: Csa), with hot and dry summers. Particularly, in August, the daily mean temperature is 25.2 °C (average high: 32.7 °C) and a RH average of 48%. The eclipse showed an appreciable impact in temperature, RH as well as in VP as it can be seen from the measurements by this author of the variables involved. The author provided seventeen readings of each between 13:05 and 14:53 near ground level. Clayton (1908) in his figures. 5, 11 and 15 of Plate I of his paper published plots made with these Jaubert's data of temperature, VP and RH, respectively. In particular, the data we give in Table 3 from Jaubert (1906) are at the time of 13:05, 14:12, 14:29 and 14:53. These time values do not coincide to eclipse contacts; rather, they were selected on the basis of minimum and maximum values of RH and VP, in between the first and last values. Then, within this interval (which includes totality), a dip of 4.2 °C was detected during the eclipse; RH also decreased by 4.2% and VP decreased by 6.8 hPa. The fact that RH had decreased reveals what Clayton (1908) stated (see subsection 2.2).

Given the set of measurements of VP provided by Jaubert (1906) it is worthy to check the values by applying Eq. (1). The results, not included here, show an excellent agreement indicating the accuracy with which he made them at that time of 1905.

Another eclipse observed during a summer was that of 11 August 1999, in which Hassan *et al.* (1999) made measurements of temperature and RH at Helwan, Egypt; there the eclipse was partial eclipse with 62.33% occultation. Köppen-Geiger climate classification system classifies its climate as hot desert (BWh). Owing to its proximity to Cairo, its average monthly temperatures are quite similar, but it has a quite different distribution of humidity and its diurnal average temperature variation is slightly larger. They reported at 1C a temperature value of 37.2 °C and a value of 35.9 °C at maximum partiality; the value at 4C was of 37.5 °C. Also, RH values, for the same instants, were reported: 18%, 21% and 18%, respectively. Vapor pressure values, calculated via Eq. (1), show an increase from 11.42 hPa at 1C to 12.41 hPa at maximum partiality.

Finally, at the end of Table III, we have the partial eclipse of 3 October 2005 observed in Hada Al-Sham area, Makkah (Mecca), Saudi Arabia, by Anbar (2006). With an occultation of 50.33% this eclipse produced no impact on temperature but certain impact in RH and VP as shown in this table. The first two variables were measured at heights of 3.5 m and 5.5 m above ground level (agl). The RH yielded the same value for both heights. The values shown in the table were taken out from plots published by this author. The impact on temperature was pretty small (with a barely dip of 0.2 °C for both heights). The RH showed an oscillation from 63.8% (at 1C) to 66.7% (at 4C) with a minimum value of 55.7% at 12:51.4 (before maximum eclipse). By Eq. (1) we were able to estimate values for VP. At 3.5 m it oscillated from 47.8 hPa (at 1C) to 42.6 hPa (at 4C) with a minimum of 39.0 hPa (before maximum eclipse); at 5.5 m something similar occurred, it oscillated from 44.1 hPa (at 1C) to 45.0 hPa (at 4C) with a minimum of 39.0 (before maximum eclipse). There was a remarkable increase in the amount of aqueous vapor, which commenced close to but before the middle of the eclipse. This oscillation occurred at the two heights involved with values at 3.5 m greater than those at 5.5 m. in accordance with the fact that VP decreases with height. It seems, then, that change in humidity during the eclipse was most probably because the humidity is more sensitive to it than air temperature. The author emphasizes that wind speed was not to strong (it never exceeded the value of 5 m-s$^{-1}$), allowing for seeing difference in RH (and hence in VP); otherwise, higher wind could remove the effect on humidity.

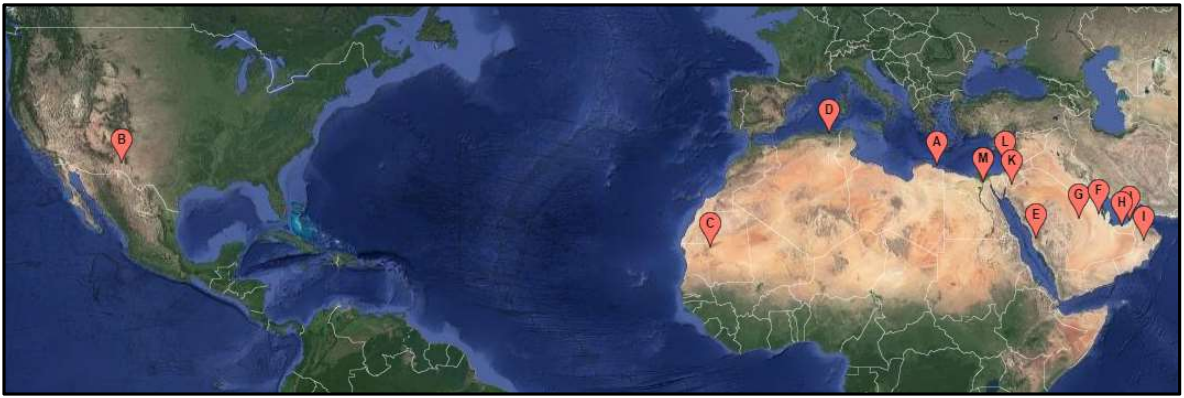

**Figure. 16**. Desert and similar places where solar eclipses have been observed, to measure air temperature and/or humidity response, between 1905 and 2020. (A) Tobruk, Lybia (2006); (B) Tularosa Basin, New Mexico, U.S.A. (1994); (C) Chinguetti, Mauritania (1973); (D) Constantine, Algeria (1905); (E) Makkah, Hada Al-Sham, Saudi Arabia (2005); (F) Al-Hofuf, Saudi Arabia (2019); (G) Riyadh, Saudi Arabia (2020); (H) Station #1, United Arab Emirates (2019); (I) Station #2, Oman (2019); (J) MWR (close to Dubhai's International Airport), United Arab Emirates (2019); (K) Eilath, Israel (1952); (L) Tel Aviv, Israel (1952); (M) Helwan, Egypt (1999).

**Table III.** Solar eclipse temperature and humidity observations in desert and similar environments in the Near East, northern Africa and in the U.S.A. between 1905 and 2006

| Eclipse date, site, occultation, duration & reference | Contacts (C) (Local time) [hr:min:s] | Temperature (at time) [°C] or Temperature change | RH (at time) [%] | VP (at time) [hPa] |
|---|---|---|---|---|
| 25 February 1952 Eilath, **Israel** (29°33'.4N, 35°56'9E, 63 m asl) Partial: 75.30% 2 hr 42 min 48 s Klein & Robinson (1955) | 1C / 10:17:06 --- Max / 11:39:10 --- 4C / 12:59:54 | ~18.3 (1C) --- ~17.4 (Max) --- ~ 19.1 (4C) | ~ 47.8 (1C) --- ~ 46.7 (Max) --- ~ 48.9 (4C) | ~ 9.4 (1C) --- ~9.9 (Max) --- ~9.2 (4C) |
| 25 February 1952 Tel Aviv, **Israel** (32°4'.85N, 34°46'.8E, 15 m asl) Partial: 69.19% 2 hr 39 min 11 s Klein & Robinson (1955) | 1C / 10:22:03 --- Max / 11:42:20 --- 4C / 13:01:14 | ~ 15.8 (1C) --- ~ 15.3 (Max) --- ~ 16.8 (4C) | - | - |
| 29 March 2006 Tobruq, **Lybia** (32°05'N, 23°59'E, 30 m asl) Partial: 99.98% 2 hr 39 min 24 s Hassan & Rahoma (2010) | 1C / 11:19:24 --- Max / 12:39:01 --- 4C / 13:58:48 | 19.5 (1C) 18.0 (Max) 21.0 (4C) | 62 (1C) 82 (Max) 78 (4C | [via Eq. (1)] 14.04 (1C) 16.91 (Max) 19.38 (4C |
| 10 May 1994 Tularosa Basin, **New Mexico, U.S.A.** (32°24'N, 106°21'W, 1220 m asl) Annular: 86.65% 3 hr 08 min 19 s Eaton *et al.* (1997) | 1C / 08:46:41 --- 2C / 10:10:51 --- 3C / 10:15:13 --- 4C / 11:55:00 | "the near-surface air was stable, form the surface to 20 m AGL, for about 45 min centered around the time of maximum eclipse. Air temperature at the 4 m level decreased about 3 °C during the eclipse…" | | |
| 30 June 1973 Chinguetti, **Mauritania** (20°26'.4N, 12°15'.7W, 453 m asl) Total: 100% 2 hr 50 min 16 s Anderson & Keefer (1975) | 1C / 09:28:17 --- 2C / 10:45:41 --- 3C / 10:51:58 --- 4C / 12:18:33 | The following absolute anomalies (Tmax-Tmin) were found by the authors: "The temperature changes were 3.5°C at 0.3 m and 2.5°C at 6.75 m and 13.5 m above the surface". | - | - |
| 30 August 1905 Constantine, **Algeria** (36°21'.9N, 6°36'.87E, 574 m asl) total: 100% 2 hr 37 min 23 s Jaubert (1906) | 1C / 12:21:45 --- 2C / 13:42:17 --- 3C / 13:45:27 --- 4C / 14:59:08 | 32.6 (13:05) 29.4 (14:12) 28.4 (14:29) 30.2 (14:53) | 20 (13:05) 19 (14:12) 43 (14:29) 30 (14:53) | 9.83 (13:05) 7.78 (14:12) 16.62 (14:29) 12.87 (14:53) |
| 11 August 1999 Helwan, **Egypt** 29°50'.99N, 31°21'.0W, 28 m asl Partial: 62.33% 2 hr 46 min 41 s Hassan *et al.* (1999) | 1C / 13:10:44.6 --- Max / 14:38:27.8 --- 4C / 15:57:26.2 | 37.2 (1C) --- 35.9 (Max) --- 37.5 (4C) | 18 (1C) --- 21 (Max) --- 18 (4C) | [via Eq. (1)] 11.42 (1C) --- 12.41 (Max) --- 11.61 (4C) |

| | | | | [via Eq. (1)] |
|---|---|---|---|---|
| 3 October 2005 Makkah, Hada Al-Sham **Saudi Arabia** (21°48'.1N, 39°43'.7E, 245 m asl) Partial: 50.33% 3 hr 06 min 28 s Anbar (2006) | 1C / 11:58:56 --- Max / 13:34:17 --- 4C / 15:05:24 | ~ *40.10* (3.5 m agl) ~ *38.76* (5.5 m agl) --- **~ 40.25** (3.5 m agl) **~ 39.90** (5.5 m agl) --- ~39.5 (3.5 m agl) ~38.5 (5.5 m agl) | ~ *63.8 (11:59.0)* --- **~ 55.7 (12:51.4)** --- ~66.1 (15:05.4) | ~ *47.3 (3.5 m agl)* ~ *44.0 (5.5 m agl)* --- **~ 41.6** (3.5 m agl) **~ 40.9**(5.5 m agl) --- ~46.2 (3.5 m agl) ~45.0 (5.5 m agl) |

Notes. Approximated values taken from the plots published in the indicated references are leveled with the symbol "~". Heights with "agl" means "above ground level" and with "asl" means "above sea level".

## 6.2. Present-day observations

The recent annular solar eclipse in the early morning of 26 December 2019, in the Arabian Peninsula, aroused interest in other different places for its observation (see Figure 16). For example, observations in the United Arab Emirates and Oman were made whose results and analysis have already been published by Nelli *et al*. (2020) for three observation sites, which the authors call "station #1" [Figure 16 (H)] on the center line (at 23° 30' N, 53° 30' E, ~ 100 m asl), "station #2" [Figure 16 (I)] on the center line (at 21° 30' N, 57° 00' E, ~ 100 m asl) and "MWR" [Figure 16 (J)] (24° 26' 11" N, 54° 36' 43" E) about 4 km from Abu Dhabi's International Airport. Table IV sums up the changes in temperature in the sites considered by Nelli *et al*. (2020) noting that no appreciable change was observed in station #1 like that found for Riyadh in this work. Nelli *et al*. (2020) discuss this outcome on the base of clearer and drier conditions compared with adjacent days, which enhanced radiative cooling and so producing lower temperatures. At both sites the eclipse was not enough to produce significant change. However, at stations #2 and MWR changes were detected. The change of 6°C in station #2, a little later, is striking when it is compared with other results from previous studies cited by Nelli *et al*. (2020) in which changes were lower; and it is even more striking if we consider that this eclipse was early in the morning. There is no doubt that desert conditions are more influential over the rest of others (excepting those in the Arctic or Antarctic). On this respect and as a reference, compare in Table III, for example, the response between Tel Aviv and Eilath during the partial eclipse of 25 February 1952 in Israel.

**Table IV**. Temperature and humidity changes during the solar eclipses of 26 December 2019 and of 21 June 2020 in the Arabian Peninsula

| Eclipse date, site, occultation, duration & reference | Contacts (C) (Local time) [hr:min:s] | Temperature change or Temperature (at time) [°C] | RH (at time) [%] | VP (at time) [hPa] |
|---|---|---|---|---|
| 26 December 2019 Station #1 United Arab Emirates 23° 30' N, 53° 30' E, ~100 m asl Annular: 91.738% Nelli et al. 2020 | 1C / 06:31:14.7 — 2C / 07:35:06.7 — Mid / 07:36:36.7 — 3C / 07:38:06.7 — 4C / 08:51:55.8 | "Consequently, LST signal from the eclipse is not evident at location #1, even though it may still be present…" | - | - |
| 26 December 2019 Station #2, Oman 21° 30' N, 57° 00' E, ~100 m asl Annular: 91.947% Nelli et al. 2020 | 1C / 06:30:21.8 — 2C / 07:36:16.1 — Mid / 07:37:36.6 — 3C / 07:38:57.0 — 4C / 08:55:49.4 | "At station #2, the largest temperature difference was about 6 °C and it occurred just after the ASE…" | - | - |
| 26 December 2019 MWR (~Dubhai's Int. Airport) United Arab Emirates 24°26'11"N, 53°36'43" E, ~0 m asl Partial: 90.79% Nelli et al. 2020 | 1C / 06:31:28.9 — Mid / 07:37:23.6 — 4 C / 08:53:25.6 | "After the sunrise at ~03 UTC, the surface gradually warmed up to 14.7 °C, but then it cooled down to 13.4 °C just before 04 UTC…" | - | |
| 26 December 2019 Al-Hofuf, Saudi Arabia (this work, see Table I) | See Table I | 9.4 (sunrise) 9.4 (2CD) 9.4 (3CD) 11.1 (4CD) | 96.55 (sunrise) 96.63 (2CD) 96.55 (3CD) 95.37 (4CD) | [via Eq. (1)] 11.38 (sunrise) 11.39 (2CD) 11.38 (3CD) 12.55 (4CD) |
| 21 June 2020 Riyadh, Saudi Arabia (this work, see Table I) | See Table I | 33.42 (1CJ) 35.93 (07:39:36) 32.71 (08:39:36) 37.94 (4CJ) | 19.79 (1CJ) 17.63 (07:39:36) 17.20 (08:39:36) 14.95 (4CJ) | [via Eq. (1)] 10.19 (1CJ) 10.43 (07:39:36) 8.51 (08:39:36) 9.87 (4CJ) |

Notes. MWR: microwave radiometer; LST: land surface temperature; ASE: annular solar eclipse; CD: December eclipse contacts (Table I); CJ: June eclipse contacts (Table I).

In Figure 17, with data taken from Tables III and IV, we compare graphically the VP response for seven different eclipses at different time of the day. We note a decrease in this variable for four of them, including the total eclipse at Constantine where an oscillation was detected; this is in agreement with what it is expected. In the very early morning eclipse at Al-Hofuf there was no response as it was within the natural variability at that time. Conversely, at Helwan there was an increase.

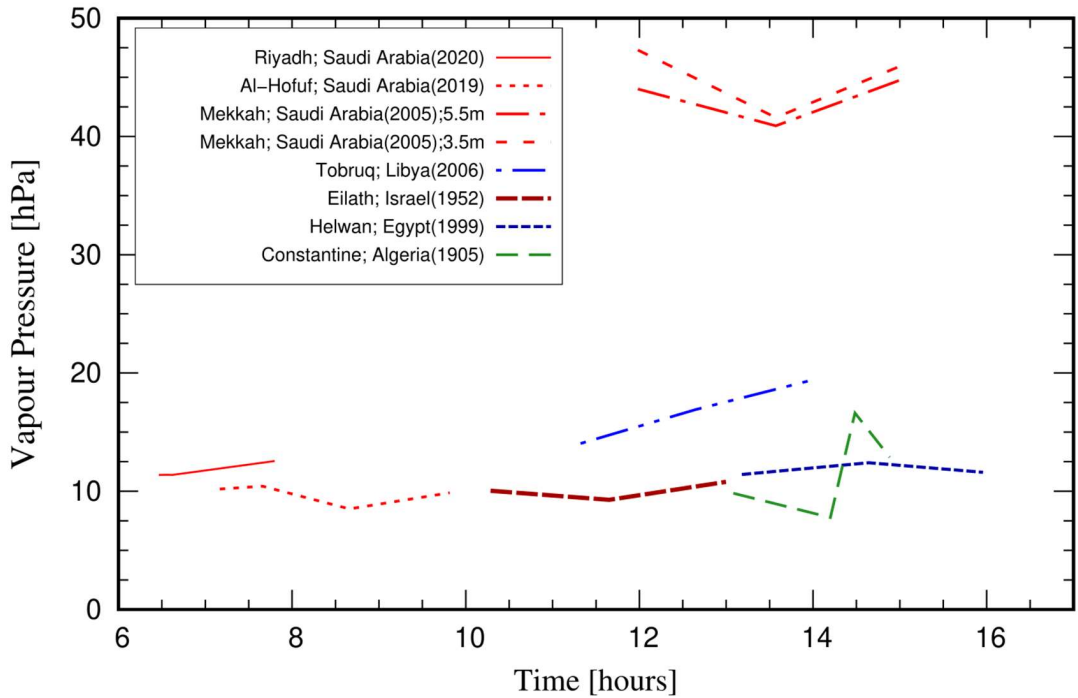

**Figure 17**. Response of VP during partial solar eclipses (except for Constantine, Algeria, where the eclipse was total) in desert or similar environments at different time of the day. In Riyadh, Eilath, Mekkah and Constantine there was a decrease in VP regardless the time of the eclipse. At Mekkah VP was estimated at two heights (3.5 m and 5.5 m agl) with the highest values (between 40 and 50 hPa). At Constantine, where the eclipse was total, an oscillation was detected. During the very early morning eclipse of Al-Hofuf no response was found. In Tobruq and Helwan the response was quiet dissimilar (all data taken from Tables III and IV).

## 7. Discussion

Regarding annular solar eclipses as astronomical phenomena that affect micrometeorological variables such as temperature, humidity, solar radiation on Earth, this article has been aimed to investigate the main anomalies produced in the temperature and humidity (RH and VP) due to the recent two Saudi Arabia solar eclipses of 26 December 2019 (annular class at Al-Hofuf City), and of 21 June 2020 (partial class at Riyadh City), which occurred in different seasons (6 months apart). The two solar eclipses were captured in progress as shown in the images of Figure 2. While the December eclipse data was a 10-second time cadence, the measurements of the June eclipse were sampled at 1-minute intervals during the eclipse phases. This allowed for an evaluation of how the variables involved responded to solar eclipses in a detailed manner. Interestingly, both eclipses were morning events under clear-sky arid fair areas, but with different situations and circumstances, say, June's eclipse starting after sunrise and December's eclipse before sunrise. Certainly, this difference impacted the temperature and RH data in different manners.

The main results found in the analysis can be summarized as follow:

- For December´s eclipse, a comparison with non-eclipse day eclipse data evidenced a change in both temperature and RH [Figure 9 - (a) and (b)], that we quantified to be approximately a 1°C drop in temperature and an increase in RH of about 3%. This impact is conspicuously less severe compared to what we observed for the June 21 eclipse where we report a clear drop in temperature as high as 3.2ºC, from a value of 35.9ºC (at 07:39 am) to 32.7ºC (at 08:35 am). In this eclipse RH values displayed a decrease trend from C1J (see Table I) to right before maximum; then, a partial increase trend was noted up to 08:41:36 from where a decrease trend was again observed, all of that absent on the non-eclipse day. This oscillation in the behavior of RH is sometime observed in eclipses when, as explained by Clayton (1908), the two opposite effects controlling this variable are so nearly equal (see section 2.2). Consequently, a similar oscillation is exhibited in VP for the June 21's eclipse as shown in Figure 14.

- In relation to time lag (from maximum occurrence time), we measured a value of about 15 min from June eclipse's temperature curve. This delay is in agreement with previous results reported in literature. It turned out that this variable was typically between 5 and 30 min post mid-eclipse, and that interestingly it might be linearly related to the global solar radiation at the end of the eclipse events, i.e., 4th contacts, (Anderson, 1999; Clark, 2016; Kameda *et al*., 2009).

- Aside from the differences in magnitudes and general trend of the meteorological data for the two eclipses, it is important to evaluate closely the related-dynamical variations. Thus, we proceeded to evaluating the data rates. The rate calculations imply estimating first derivatives of the studied parameters variables, and subsequently a "moving-average" smoothing technique has been applied for a better curve-evolution visualization. Our results are depicted for Al-Hofuf eclipse in Figure 15 (a) – (b), and for Riyadh in Figure 15 (c) – (d). Furthermore, we calculated rates for non-eclipse days (blue short-dashed curves) for comparison. Horizontal dotted lines indicate the zero-rate levels for a better interpretation of these variations. The two eclipse-related rate-curves are evidently different in various aspects. Firstly, for December 2019's eclipse, under sunrise condition, but just before the annular phase, the effects are rather less severe or unnoticed compared to June 2020's eclipse. On the one hand, we expect that after sunrise temperature increases, which can be seen from both curves (eclipse and non-eclipse days) increasing at positive rates [Figure 15 (a)]. In contrast, the effect of the eclipse is seen as a decrease of the observed trend compared to non-eclipse day, which is distinguishable just at the annularity and, later on, as a distinct difference between the two curves (dashed and continuous lines). On average, the non-eclipse day related-curve is above of the eclipse related-curve by a difference of about 0.022 °C/minute (~1.32 °C/hr) is measured between the two curves, which lasted almost until the end of the eclipse. Outside the eclipse phase the two rate curves are very similar in both trend and magnitude.

- Similar to what we have noted in the temperature, RH rates are very similar (although subjected to some additional fluctuations) for both days. After sunrise, we visualize RH negative rates (i.e., decreasing), which we certainly observe in our rate curves. A difference to be noted, however, is that given the eclipse effects RH for the eclipse day should decrease slowly compared to the non-eclipse day; this is appreciated in our rate curves where overall larger values in the continuous-curve are compared to the dashed-curve up to the end of the eclipse phase [Figure 15 (b)].

- For June's eclipse, there is a temperature increasing rate trend after sunrise [Figure 15 (c)]. About 10 min after the beginning of the eclipse phase, the computed values reveal a clear change in the rates, with an evident decrease from a value of ~ 0.078 °C/min to a negative rate of ~ -0.075°C/min in about 20 min. Subsequently, the rates increased retaining its negative values; afterwards, when the temperature reached the dip (with its characteristic lag) the rate-curve crossed again the zero value (no-dynamics), returning hereafter to an increasing temperature phase, characterized by a positive rate, reaching values as high as ~ 0.1°C/min, and maintaining positive rates for the rest of the eclipse time. Interestingly, at similar phase or time interval for the non-eclipse day, the rates are almost stable with positive values (i.e., a smooth increasing temperature trend).

- As for RH, after sunrise a negative trend in the rates (though very weak) is displayed. It continues decreasing at different rates until the curve crosses the zero-level a few min just before maximum where, it switches to positive rates indicating an increase in RH values after the maximum phase. Finally, about 25 min past maximum, negative rates even with small values continued implying a decrease in RH. These dynamics, once more, are absent in the non-eclipse day data (blue dashed line), with negative and almost stable rates all over the eclipse phase, due to post sunrise effects [Figure 15 (d)].

- With respect to the inter-comparison made among eclipses observed in desert or similar environments, and in particular those between summer total eclipses featured in Table III (Mauritania and Algeria), RH and VP presented the typical oscillation referred to above for Riyadh eclipse. In this sense, it is opportune to cite Clayton (1908) who, to explain this effect, cites in turn the paper of Eliot (1899) who, observing the TSE of 22 January 1898, in India at Nagpur (Maharastra), wrote following: "There was a remarkable increase in the amount of aqueous vapor, which commenced about the middle of the eclipse, and which was followed by an equally rapid decrease. This oscillation occurred at all stations almost without exception during the second-half of the eclipse. The period averaged half an hour and the amplitude was from 20 to 50 per cent of the mean aqueous vapor of the day… The oscillation was not due to actual horizontal air movement, but to some wave-like action transmitted very rapidly from west to east, an action similar in its rapidity to the march of the solar eclipse across India (in about 35 min) … The only action which would give rise to this large change is the descent of air masses containing a larger quantity of aqueous vapor than the air in the lower stratum displaced by this descent." Clayton (1908) adds that there was, however, one factor acting on the VP overlooked by Eliot (1898), and that was the calm which followed the eclipse shadow: "In the case of unventilated thermometers, the increase of vapor tension may be, in part at least, only apparent and due to lack of ventilation. For this reason, all eclipse observations ought to be made with aspirated or whirled psychrometers." It must be bear in mind that Nagpur in January has a dry environment with an average RH of 54%, with a maxima/minima of 29°C/13°C, and an average rain of 1 day. Therefore, it is highly probable that this mechanism has been the responsible for the oscillation seen in VP in the Chinguetti's June total eclipse of 1973 (Mauritania), the Hada Al-Sham's October partial eclipse of 2005 (Saudi Arabia), and the Riyadh's June partial eclipse of 2020 (Saudi Arabia).

## 8. Conclusions and final comments

The present investigation and the data reported and analyzed here will give important insights for future worldwide large efforts in meteorological-related research. Moreover, the data and finding can be used for validation of model predictions and forecasting of meteorological variables response to possible changes during future solar eclipses, especially with similar geographical and environmental conditions.

In contrast, similar inter-comparison studies can be suggested and conducted for eclipses occurring under other extreme environments like, for example, near polar regions or in them by using the few works that have been published. In Pasachoff *et al*. (2016), during the TSE of 20 March 2015 in Longyearbyen, Svalbard Archipelago (Norway), we find that temperature response was almost inappreciable as it was found by Maturilli & Ritter (2016) during the same eclipse and at the same archipelago (similar results from Riyadh and station #1 in the U.A.E. during the solar eclipse of 26 December 2019). However, in an even extreme cold and dry environment, we see that the response was different and more remarkable during the TSE of 23 November 2003, in Antarctica (Kameda *et al*., 2009; Klekociuk, 204). In near-polar environments like that of West Penn Island on the shore of Hudson Bay, Canada, Stewart & Rouse (1974) found an abrupt and surprising fall in temperature of 10 °C during the 90% partial eclipse of 10 July 1972 [an approximated fall value of 6 °C was found by Nelli *et al*. (2020) during the 26 December 2019 in station #2 of the U.A.E.]. Also in Svalbard a previous 93% partial solar eclipse took place on 1 August 2008; in that opportunity Sjöblom (2010) reported that the atmospheric response was much slower and weaker over water than over land, being the air temperature change of 0.3°C – 1.5°C, depending on the condition under which measurements were made at five different observing stations, etc. It seems that the maximum temperature decrease reported by Stewart & Rouse (1974) is the largest ever measured during a solar eclipse [see table 1 of Kameda *et al*. (2009)]. For those works on eclipse meteorology in Arctic/Antarctic environments or sites nearby a table, like our Table III or Table IV in this work, could be constructed in order to inter-compare its results.

It is expected that the impact on atmospheric response by solar eclipse at sunrise be imperceptible or a last minimal as occurred during the solar eclipse of 26 December 2019 in Al-Hofuf or at station #1 in the U.A.E. At sunset this result is also expected as occurred during the 99.2% partial eclipse of 11 August 1999 in a semi-arid region of India located in Ahmedabad, when a barely decrease of ~0.50 °C was measured between 16:55:29 and 18:58:48 (Indian eastern time), with maximum phase at 18:00:28 and sunset at 19:25 (Krishnan *et al*., 2004).  This time the RH did not show any appreciable variation compared to that on non-eclipse days. Similar results were obtained by the Williams College expedition to New Mexico (U.S.A.) to observe the annular solar eclipse of 20 May 2012 close to sunset. Conversely, the impact of the December 2019 solar eclipse was quite different in other places beyond the Arab world.  In a semi-arid region of southern India, called Anantapur, Reddy T *et al*. (2020) found that during eclipse day, the near-surface air temperature was significantly decreased by an amount of 1.12 °C, and RH increased 14.63% after 60 min of the beginning of the eclipse. This eclipse began at Anantapur at 08:06 IST, reached a maximum occultation of the Sun of 84.25% at 09:38 IST and ended at 11:10 IST. The total duration of the eclipse was of 3 hours and 4 min between 08:06 IST and 11:10 IST.

The results summarized in this paper, including those from other works (reviewed in sub-sections 2.2 and 2.3, and inter-compared in section 6) reveal the potential of eclipse meteorology for increasing our knowledge about the response of air temperature and humidity to solar eclipses in desert and other similar environments. The value of this information will depend on what will have already been learned through these investigations.

This study constitutes an important contribution to study eclipse-induced atmospheric changes in desert environments like that of the Arabian Peninsula**,** as part of an international eclipse meteorology collaboration currently in development in the Astronomy Department of Williams College (Williamstown, Massachusetts, U.S.A.) The international eclipse meteorology team of this institution has been collecting and analyzing atmospheric data during several past solar eclipses around the world (partial, annular and total) producing thus an important database and results (Peñaloza-Murillo & Pasachoff, 2015; Pasachoff *et al.*, 2016; Peñaloza-Murillo & Pasachoff, 2018; Peñaloza-Murillo *et al.*, 2020).

Acknowledgements

J.M. Pasachoff's eclipse research at Williams College receives major support from grant AGS-1903500 from the Solar Terrestrial Program, Atmospheric and Geospace Sciences Division, U.S. National Science Foundation. For the 2019 total eclipse, we had additional student support from NASA's Massachusetts Space Grant Consortium; Sigma Xi; the Global Initiatives Fund at Williams College; and the University of Pennsylvania. We thank the Dean of Faculty at Williams College for the support to M. A. Peñaloza-Murillo at Williams College in 2018-19, which followed his earlier Fulbright fellowship there. The research work of A. Elmhamdi in this project was supported by King Saud University's Deanship of Scientific Research and College of Science Research Center in Saudi Arabia. The bibliographical assistance of Helena Warburg of the Schow Science Library of Williams College is highly appreciated. Also, we acknowledge the permission from Narendra Reddy Nelli, Jay Anderson and Michael Zeiler to use images shown in this paper.